%% file: sample-manuscript.tex
\documentclass[acmsmall,screen]{acmart}

\usepackage[utf8]{inputenc}
\usepackage[T1]{fontenc}
\usepackage{microtype}

\AtBeginDocument{%
  }

\usepackage{custom}
\usepackage{diagbox} 

\usepackage{graphicx}
\usepackage{xcolor}
\usepackage{tcolorbox}
\usepackage{tabularx}
\usepackage{multirow}
\usepackage{amsmath,amsfonts}

\usepackage{textcomp}
\usepackage{tcolorbox}
\usepackage{colortbl}
\usepackage{algorithm2e}
\usepackage{listings}
\AtBeginDocument{\DeclareCaptionSubType{lstlisting}}
\usepackage{caption,subcaption}

\definecolor{codegreen}{rgb}{0,0.6,0}
\definecolor{codegray}{rgb}{0.5,0.5,0.5}
\definecolor{codepurple}{rgb}{0.58,0,0.82}
\definecolor{backcolour}{rgb}{0.95,0.95,0.92}

\lstdefinestyle{mystyle}{
    backgroundcolor=\color{backcolour},   
    commentstyle=\color{codegreen},
    keywordstyle=\color{magenta},
    numberstyle=\tiny\color{codegray},
    stringstyle=\color{codepurple},
    basicstyle=\ttfamily\footnotesize,
    breakatwhitespace=false,         
    breaklines=true,                 
    captionpos=b,                    
    keepspaces=true,                 
    numbers=left,                    
    numbersep=5pt,                  
    showspaces=false,                
    showstringspaces=false,
    showtabs=false,                  
    tabsize=2,
    xleftmargin=1em
}

\setcopyright{rightsretained}
\acmDOI{10.1145/3660809}
\acmYear{2024}
\copyrightyear{2024}
\acmSubmissionID{fse24main-p865-p}
\acmJournal{PACMSE}
\acmVolume{1}
\acmNumber{FSE}
\acmArticle{102}
\acmMonth{7}
\received{2023-09-28}
\received[accepted]{2024-04-16}

\begin{document}

\title{Mining Action Rules for Defect Reduction Planning}

\author{Khouloud Oueslati}
\orcid{0009-0002-9641-885X}
\affiliation{%
  \institution{Polytechnique Montréal}
  \city{Montreal}
  \country{Canada}
}
\email{KHOULOUD.OUESLATI@POLYMTL.CA}

\author{Gabriel Laberge}
\orcid{0009-0002-2019-2292}
\affiliation{%
  \institution{Polytechnique Montréal}
  \city{Montreal}
  \country{Canada}
}
\email{gabriel.laberge@polymtl.ca}

\author{Maxime Lamothe}
\orcid{0000-0003-3705-6238}
\affiliation{%
  \institution{Polytechnique Montréal}
  \city{Montreal}
  \country{Canada}
}
\email{maxime.lamothe@polymtl.ca}

\author{Foutse Khomh}
\orcid{0000-0002-5704-4173}
\affiliation{%
  \institution{Polytechnique Montréal}
  \city{Montreal}
  \country{Canada}
}
\email{foutse.khomh@polymtl.ca}


\begin{abstract}
Defect reduction planning plays a vital role in enhancing software quality and minimizing software maintenance costs. By training a black box machine learning model and “explaining” its predictions, explainable AI for software engineering aims to identify the code characteristics that impact maintenance risks. However, post-hoc explanations do not always faithfully reflect what the original model computes. In this paper, we introduce CounterACT, a Counterfactual ACTion rule mining approach that can generate defect reduction plans without black-box models. By leveraging action rules, CounterACT provides a course of action that can be considered as a counterfactual explanation for the class (e.g., buggy or not buggy) assigned to a piece of code. We compare the effectiveness of CounterACT with the original action rule mining algorithm and six established defect reduction approaches on 9 software projects. Our evaluation is based on (a) overlap scores between proposed code changes and actual developer modifications; (b) improvement scores in future releases; and (c) the precision, recall, and F1-score of the plans. Our results show that, compared to competing approaches, CounterACT’s explainable plans achieve higher overlap scores at the release level (median 95\%) and commit level (median 85.97\%), and they offer better trade-off between precision and recall (median F1-score 88.12\%). Finally, we venture beyond planning and explore leveraging Large Language models (LLM) for generating code edits from our generated plans. Our results show that suggested LLM code edits supported by our plans are actionable and are more likely to pass relevant test cases than vanilla LLM code recommendations.

\end{abstract}






\maketitle

\input{sections/0_intro}
\input{sections/1_background}

\input{sections/Action_rule_mining}
\input{sections/2_methodology}

\input{sections/3_results}
\input{sections/4_discussion}

\input{sections/5_threatsToValidity}
\input{sections/conclusion}

\bibliographystyle{ACM-Reference-Format}
\bibliography{biblio}

\end{document}

%% file: sections/0_intro.tex
\section{Introduction}
In today’s software-driven society, Software Quality Assurance (SQA) has become a crucial practice to ensure that software products meet the required standards of functionality, reliability, usability, and performance. Since fixing defects is costly~\cite{hawking1988}, addressing them before the software is released can typically result in faster and cheaper remediation~\cite{Boehm}, yet it remains a critical challenge to identify and fix defects early in the development lifecycle. Over the past decade, numerous Software Defect Prediction (SDP) techniques emerged in order to predict the high-risk areas of source code that are prone to post-release defects~\cite{8263202, shin2021explainable,Menzies121,NagappanQWQ,AmbrosAS}. However, one of the significant limitations faced by SDP models is their lack of explanation and actionability. Due to the absence of practical recommendations that developers can act upon, practitioners might be skeptical of predictions that seem counterintuitive which may hinder the adoption of software analytics in practice \cite{DBLP:journals/corr/abs-1802-00603}. Indeed, a prior survey investigating practitioners’ perceptions of SQA rule-based planning~\cite{sqaplanner}, found that 52\%-80\% of respondents perceived it as useful, and 52\%-72\% of the respondents expressed willingness to adopt rule-based guidance. Furthermore, the release of commercial AI-driven defect prediction tools such as Amazon’s CodeGuru, and Microsoft’s Code Defect AI, implies practitioners’ interest in defect reduction planning. The latter even uses explainable Just-In-Time defect prediction to provide a visualization of the software metrics that contribute to general code improvement guidelines. While several research studies have leveraged model-agnostic post-hoc explainability techniques such as LIME \cite{Jiarpakdee} and TimeLIME~\cite{peng2021defect}, to understand file-level defect predictions of black-box ML models, post-hoc explanations are not always trustworthy. They can be misleading and do not always faithfully reflect what the original model computes~\cite{rudin2019stop}. Moreover, previous work~\cite{shin2021explainable} has shown that they can be inconsistent and unreliable under different settings \cite{DBLP:journals/corr/abs-2111-10901}. Consequently, this risks making practitioners hesitant to rely on model-agnostic techniques for taking defect-reducing actions \cite{10123486}.

In this paper, in contrast to the current trend of leveraging black-box models, we show that it is possible to provide software improvement plans by mining 
action rules from historical data and filtering out those that overlap the least with past code changes.
Based on this idea, we propose CounterACT, an approach that provides counterfactual explanations based on Action Rule Mining (ARM) \cite{Sykora2020ActionRC}. By running simulations over the historical records, CounterACT generates recommendations that help in preemptively reducing the likelihood of potential defects by explaining how an action (a change in the value of one or more flexible attributes) could impact the classification of a given object (e.g., from defective code to working code). Since it is inferred from classification rules mining, the process that generates the plans is fully transparent and a developer can understand why a certain plan was recommended.

To assess the effectiveness of CounterACT, we answer the following research questions:

\textbf{RQ1: How effective is the rule-based guidance generated by CounterACT?}
At the release level, the rule-based guidance generated by our approach achieves a high True Positive rate of an average 97\%  meaning that the learned action rules highly match historical bug fixes done by developers. In addition, the action rules, even when derived from less frequent patterns (as indicated by the average support of 13\%), have a high median confidence of 89\%. This implies that when a bug fix is recommended by an action rule, there is an 89\% chance that the expected non-buggy outcome will occur. Furthermore, the positive uplift values, with an average of 43\% indicate that applying the recommended action will increase the likelihood of the desired outcome (reduction of defects) by 43\% compared to not taking that action.

\textbf{RQ2: How does CounterACT compare against competing defect reduction approaches at the release level?}
To showcase the effectiveness of counterfactual-based planning (using action rule mining) in generating plausible and actionable recommendations to reduce defects in subsequent releases, we compare CounterACT with TimeLIME and other post-hoc defect reduction planners as well as the original ARM algorithm using four performance criteria (i.e., overlap, improvement score, precision, and recall) across 9 projects. Our analysis shows that CounterACT plans have the highest median overlap of 95\% which implies that our suggested plans align more closely with actual developer changes when compared to other competing approaches. Additionally, the actions proposed by CounterACT are associated with a larger reduction in defects than other algorithms (median improvement score 91,33\%).

\textbf{RQ3: How does CounterACT compare against competing defect reduction approaches at the commit level?}
In order to ensure the practicality of our approach, we provide software quality improvement plans at the change level (Just-in-Time or JIT) allowing quicker defect remediation. 
To determine how well CounterACT performs against other defect reduction planners at a finer granularity, we reuse the performance criteria from RQ2 and evaluate its performance on 5 open-source projects at the commit level. Our findings show
that CounterACT also outperforms competing approaches at the commit level, where the overlap score of CounterACT is on average 85.97\%. This implies that our plans have the highest similarity with actual actions done by developers in reducing defects at a finer granularity, whereas the performance of the current state-of-the-art approach TimeLIME declines from 79.17\% at the release level to 38.36\% at the commit level. 

Furthermore, the plans provided by state-of-the-art techniques can be challenging to implement by developers because they are reduced in terms of code metrics. In this paper, we discuss the use of Large Language Models (LLMs) coupled with CounterACT plans to generate automated code edits for developers. Our experiments show that the combination of LLMs and planning can lead to enhanced guidance. Indeed, our primary results indicate that the LLM supported by CounterACT plans effectively addresses and resolves bugs in 28 out of 40 cases while the vanilla LLMs recommendations produced incoherent code edits in 26 out of 40 cases.

Overall, our findings suggest that CounterACT is a promising approach for generating plausible and 
interpretable defect reduction plans, 
outperforming existing approaches while providing actionable insights for developers. Our contributions are the following:
\begin{itemize}

    \item We introduce a novel approach CounterACT which leverages action rule mining to yield defect reduction plans.
    \item We show that CounterACT suggests plans that have higher similarity with developers' actions in reducing defects than TimeLIME's while being fully interpretable.
    \item We extend the methodology of defect reduction planning by also considering commit-level defect reductions in addition to release-level which was previously studied. 
    \item We use a Large Language Model (CodeLlama~\cite{rozière2023code}) to automate the recommendation of code changes using the plans provided by our approach, as an additional dimension to evaluate the quality of CounterACT’s commit-level recommendations.
    
\end{itemize}

%% file: sections/1_background.tex
\section{Background and related work}\label{sec:background}

\subsection{Software Defect Prediction}

\subsubsection{Machine Learning Techniques}

As software projects develop and evolve, defects are introduced, resulting in exponential costs that consume limited software 
quality budgets and resources. Over the years, numerous approaches and techniques have been proposed to proactively identify and 
mitigate defects in software systems \cite{inproceedings1,inproceedings2,10.1109/ICSE.2007.66}. Recent advancements have focused on the use of black-box ML models due to their higher predictive accuracy, SQA teams can 
use SDP models which are trained using historical data to predict the likelihood of defect-prone software 
modules in the future in order to support decisions about resource allocation in SQA activities \cite{6035727,4027145}. Indeed, SDP models were explored at 
different levels of granularity such as release level and at change level (just-in-time) \cite{10.1145/2950290.2950353,article3}. 
\subsubsection{Post-hoc Explanations}

As ML models become more complex and achieve better accuracies, understanding their decision-making process becomes 
much harder and leads to a lack of trust in the predictions. Over the years, many studies used model-agnostic techniques such as Local Interpretability Model-Agnostic Explanations (LIME)~\cite{ribeiro2016should} and Shapley Additive Explanations (SHAP) \cite{DBLP:journals/corr/LundbergL17}, to support practitioners in decision-making regarding software defects. Recent studies used the latter in selecting the most impactful features to make the black box ML models more explainable \cite{9919490}, detecting buggy lines of codes \cite{DBLP:journals/corr/abs-2103-07068}, and providing plans to reduce future defect-introducing commits \cite{9678763}.

\subsection{Defect Reduction Planning}

\subsubsection{Planners}

Despite efforts to make defect prediction models more explainable, the lack of actionability on the generated explanations 
is a roadblock toward their wide-spread application \cite{tantithamthavorn2021actionable}. Indeed, 
even if ML models could provide explanations for their decisions, we would still need to verify that those explanations 
lead to concrete and applicable recommendations. For instance, if a certain module in a software project is flagged 
as high risk by a predictive model and a tool, such as LIME~\cite{ribeiro2016should}, identifies its risk factors 
(e.g. large LOC, large Cyclometric Complexity), a developer might still want to know  ``what plan can I follow to 
reduce the risk of defects on that file?''. 

Henceforth, we will refer to a \emph{planner} $p$ as a function that takes code metrics values $m$ and outputs a
region $p(m)=R$ of ``safe'' code metric values. We will often call the safe region $R$ the \emph{plan} since it highlights
where the code metrics should be. Hence the action recommended to the developer by the planner is to 
change the code so that the new metric values land in the plan.

Before the boom in Machine Learning techniques, defect reduction planners were often provided by
empirically chosen thresholds for code complexity 
metrics~\cite{alves2010deriving,shatnawi2010quantitative,oliveira2014extracting}. The main intuition 
behind these approaches is that large values of code complexity metrics are indicative of bad coding practices 
and poor maintainability, which leads to more defects~\cite{Wahono}. Therefore, if a code metric 
$m$ has a value beyond a threshold $\gamma$, then these defect reduction planners recommend refactoring 
the code so that the metric lies under the threshold. 
\begin{equation}
    p(m) = 
    \begin{cases}
       [0, \gamma] &\quad\text{if } m > \gamma\\
       m &\quad\text{otherwise}.
     \end{cases}
\end{equation}
The different approaches differ in how the thresholds $\gamma$ are computed. For instance, Alves planner 
\cite{alves2010deriving} computes it with a weighted Cumulative Distribution Function (CDF) while Shatnawi 
\cite{shatnawi2010quantitative} computes it with a Logistic Regression. Additionally, Oliveira \cite{oliveira2014extracting}
find values of $r$ and $\gamma$ s.t. $r\% $ of the classes have metric values $m\leq \gamma$. The values of $r$ and $\gamma$ 
are obtained via an optimization problem that trades-off idealized and realistic code practices.

These three approaches have shown certain limitations: (1) they do not take into consideration historical code changes, 
and (2) they do not employ the full arsenal of highly predictive ML techniques available today. 
Therefore, they can be inaccurate or lead to plans that are surprising to developers and impossible to 
apply in practice. 

To address these two points, two ML-based defect reduction planners have recently been proposed: 
XTREE~\cite{krishna2020prediction} and TimeLIME~\cite{peng2021defect}. 
The former fits a decision tree classifier on software defect data and provides plans that change the 
prediction of the model while acting on code metrics that tend to historically vary together.
The latter employs the post-hoc explainability technique LIME~\cite{ribeiro2016should}, to identify the code metrics 
that increase the defect risk, and derives plans from these metrics while ensuring a high overlap with past developer 
changes. TimeLIME~\cite{peng2021defect} has shown to provide better plans than previous approaches in terms of overlap 
with past changes and correlation with defect reduction. Hence it is the main baseline against which we compare our approach.

Although TimeLIME can provide maintainable defect reduction plans, its main limitation is that the pipeline which 
generates the plans is opaque. Indeed, 
using TimeLIME requires first fitting a complex ML model and then approximating it locally with a 
linear model to get a recommendation. It is therefore difficult to understand why a certain plan was suggested 
and not another. Furthermore, if all provided plans are bad, the source of the error can be hard to pinpoint since 
it can be attributed to both the black box model and the explainer. In this work, we will show \emph{transparent} 
defect reduction planners are possible, thus removing these sources of confusion.

\subsubsection{Plan Quality Assessement}

To compare different planners we need to define what is a \emph{good} plan. Ideally, the best plan would be the one that 
leads to the most defect reductions when the developers follow it. However, like previous work 
\cite{krishna2020prediction,peng2021defect}, we are performing an \emph{observational} study by investigating 
historical codes changes. We do not explicitly tell the developers to follow our plans. Thus, we cannot claim that any 
of the plans can \emph{cause} a reduction in software defects, or that they can \emph{fix} bugs. Rather, the main hypothesis
behind plans is that code metrics are proxies for good software design practices which influence the prevalence of future defects. 
Hence, by changing code metrics values following a plan, we only expect developers to improve the design of their
code which should prevent \emph{future} defects. Thus, \newline

\begin{centering}
    \textit{A good plan should overlap with past code changes and be statistically associated with defect reductions.}
\end{centering}\newline

This is the main philosophy behind the K-test~\cite{Krishna2017FromPT}. More specifically, the K-test examines code regions common 
to three consecutive software releases $t-1$, $t$, and $t+1$. This region, indexed by $i$, could be an object-oriented class, a 
function, or a file present across all releases. For each version of the software, we must also have access to $M$ code metrics
so that $m^{(i)[t]}_j$ is the $j$th code metric evaluated on the $i$th code region on version $t$. 
The K-test also assumes the existence of a quality measure of the code region evaluated at each version 
$Q^{(i)[t]} $. Importantly, if $p$ is a good planner, then
quality improvement $Q^{[t]} - Q^{[t-1]}$ should be positively correlated with the \textbf{overlap} between the 
plan $p(m^{[t-1]})$ and the actual code change $m^{[t]}$. The overlap is computed via the formula
\begin{equation}
    O(p, m^{[t-1]}, m^{[t]}) = \#\big\{\,j : m_j^{[t]} \in p\big(\,m_j^{[t-1]}\,\big)\,\big\} / M,
    \label{eq:overlap}
\end{equation}
which is the ratio of code metrics from version $t$ that land inside the plan proposed in 
version $t-1$. In Table~\ref{table:overlap_example} we show an example of an overlap score.
\input{tables/overlap_example}

Following previous studies~\cite{ Mobini, COUTO201424}, we use the Number of Defect in Previous Version
$t$  \textbf{NDPV} as the quality metric $Q^{[t]}$ and the correlation between quality improvement and overlap denoted as the \emph{improvement score}  can be 
computed:
\begin{equation}
    S_{\text{scaled}} = \frac{\sum_{i=1}^{N}\big(Q^{(i)[t]}-Q^{(i)[t-1]}\big) \times O(p, m^{(i)[t-1]}, m^{(i)[t]})}
    {\sum_{i=1}^{N}Q^{(i)[t]}-Q^{(i)[t-1]}}
    \label{eq:weighted_NDPV}
\end{equation}
where the index $i$ again represents a code region (e.g. class) and so we sum over all $N$ regions. 
 This correlation is highly positive if high overlap tends to occur when the quality increases.

To identify if the developers can implement the recommended plans by \texttt{CounterACT}, as suggested in prior work~\cite{peng2021defect} we label the generated plans into one of the following categories:
\begin{itemize}
    \item \textbf{True Positive (TP):} The plan recommends changes, which are observed later on.
    \item \textbf{True Negative (TN):} The plan recommends no changes, and no changes are observed later.
    \item \textbf{False Positive (FP):} The plan recommends changes, but they are not observed later.
    \item \textbf{False Negative (FN):} The plan recommends no changes, yet some changes are observed later.
\end{itemize}

Then we calculate the precision, recall and the F1-score, as follow:
\begin{itemize}
    \item \textbf{Precision = TP/TP+FP}: Among all the recommended changes by the defect reduction planner, how many are observed in the next release?
    \item \textbf{Recall = TP/TP+FN}: Among the changes observed in the next release, how many of them match the defect reduction planner's recommendations?
    \item \textbf{F1-score = 2x(\textbf{Precision}*\textbf{Recall})/(\textbf{Precision}+\textbf{Recall})}
\end{itemize}

Indeed, a higher precision indicates that the suggested plans are more frequently utilized, while a higher recall implies fewer unexpected changes in the next release. Consequently, effective defect-reducing plans should be associated with both high precision and high recall.

%% file: tables/overlap_example.tex
\begin{table}[t]
\caption{Example of computing overlap using the Jaccard similarity function. Plans that
match the developer actions are marked gray}
\vspace{-2mm}
\centering
\scalebox{0.9}{
\begin{tabular}{|c|c|c|c|c|}
\hline
\textbf{} & AMC & LOC & LCOM & CBO \\ \hline
Current release $t$ & 2.13 & 504 & 0.9 & 5 \\ \hline
$R$ for release $t+1$ & \cellcolor{gray!50}no change & (470,520) & \cellcolor{gray!50}(0, 0.4) & no change \\ \hline
Next release $t+1$ & \cellcolor{gray!50}2.13 & 530 & \cellcolor{gray!50}0.3 & 2 \\ \hline
is a Match? & y & n & y & n \\ \hline
\end{tabular}
}
\vspace{-2mm}
\label{table:overlap_example}
\end{table}

%% file: sections/Action_rule_mining.tex
\section{Action rule mining}
\label{sec:actionrulemining}
An action rule describes how an action (a change in the value of one or more flexible attributes) could impact the classification of a 
given object. It serves as a recommendation for a course of action to take, to increase the probability of a desired class (a  
counterfactual ``what if'' explanation for the classification).  
Action Rules mining~\cite{Sykora2020ActionRC} has two phases: (1) Classification rules mining; (2) Generation of Action Rules from a subset of discovered classification rules.

\begin{figure*}[!h]
\vspace{-2mm}
    \centering
    \resizebox{0.45\linewidth}{!}{
        \input{Tikz/illustration.tex}
    }
    \vspace{-5mm}
    \caption{How action rules are mined. First classification rules with sufficient support and confidence
    are mined using Apriori. This leads to the two rules in the red and blue regions. Then, rules
    with matching antecedents but different consequents are combined leading to the action rule with support(6\%) and confidence(52\%)
    $r=[(\text{avg\_cc}>1.4) \land (\text{cbo}>14 \rightarrow \text{cbo}\leq14)] 
    \Rightarrow [\text{bug}\rightarrow \text{no-bug}]$.}
    \label{fig:illustration}
    \vspace{-3mm}
\end{figure*}
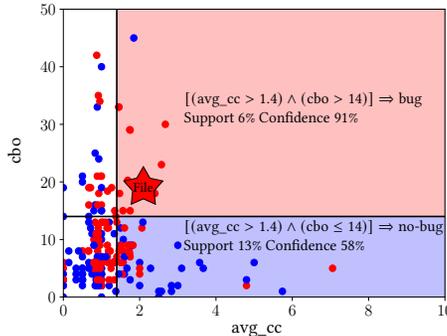

Classification rules are mined using Apriori modified for classification rule mining \cite{10.1007/3-540-45329-6_8}. They take the following form
\begin{equation}
    r_c = [(a_1 \land b_1 \land c_1 \land e_1) \Rightarrow d_1],
\label{eq:rule}
\end{equation}
where the \emph{antecedent} $a_1$, $b_1$, $c_1$ and $e_1$ represent feature values and the
\emph{consequent} $d_1$ denotes the value of the target variable predicted.
Each identified rule comes with support and confidence values, indicating its quality. 
First, support $\sup(\ante\Rightarrow \conseq)$ represents the ratio of data instances 
matching both the rule's antecedent and consequent. Secondly, confidence 
$\conf(\ante\Rightarrow \conseq)=\frac{\sup(\ante\Rightarrow \conseq)}{\sup(\ante)}$ 
is the ratio of instances fulfilling both antecedent and consequent to those satisfying only 
the antecedent. See Figure \ref{fig:illustration} for an example. We have two classification rules
that predict different outcomes: bug and no-bug. Looking at the rule 
$[(\text{avg\_cc}>1.4) \land (\text{cbo}>14)] \Rightarrow \text{bug}$, the support is $6\%$ meaning that $6\%$ 
of the data points have the specified metric values and are also buggy. The confidence is very high (91\%) meaning that, 
out of all the instances that respect the antecedent, a majority of them are indeed buggy.

Rules are interpretable by design but they describe \emph{what is}, and not \emph{what ought to be}. 
This is where action rules come into play. For example, To build an action rule, two rules with matching antecedents but different
consequents are combined:
\begin{equation}
    \begin{aligned}
    \text{(First classification rule)} \,\,\,\, &r_1 = [\omega \land \alpha] \Rightarrow \text{bug}\\
    \text{(Second classification Rule)}\,\,\,\, &r_2 = [\omega \land \beta] \Rightarrow \text{no-bug}\\
    \text{(Action Rule)\,\,\,\,} &r\,\, = [\omega \land (\alpha \to \beta)] \Rightarrow [\text{bug} \to \text{no-bug}].
    \end{aligned}
\label{eq:action_rule}
\end{equation}
In this equation, $\omega$ represents a constant condition (stable attribute), with 
$(\alpha \to \beta)$ being the suggested change to a subset of flexible attributes. 
(\text{bug} $\to$ \text{no-bug}) symbolizes the desired change in terms of prediction.
Going back to the example of Figure \ref{fig:illustration}, the two rules presented can be 
combined into the action rule  $a=[(\text{avg\_cc}>1.4) \land (\text{cbo}>14 \rightarrow \text{cbo}\leq14)] 
\Rightarrow [\text{bug}\rightarrow \text{no-bug}]$ which suggests diminishing the Coupling Between
Objects (CBO) of the current class. The metrics used to assess the quality of an action rule differ from those of a classification rules. Indeed, the support is defined as the minimum support between the 
two rules combined:
\begin{equation}
    \supp(r) = \min\{\supp(r_1), \supp(r_2)\},
    \label{eq:supp}
\end{equation}
while the confidence becomes the product of the two confidences:
\begin{equation}
    \conf(r) = \conf(r_1)\times \conf(r_2).
    \label{eq:conf}
\end{equation}
According to literature \cite{ActionRuleMining}, the intuition behind this new definition of confidence is as follows: consider all pairings between 
examples that respect the antecedent of $r_1$ and those that respect the antecedent of $r_2$. Then, 
compute the number of pairs where the label switches from bug to no-bug. This ratio is the confidence of the action rule. Lastly, the uplift \cite{Kumar2018UpliftM} of an action rule is defined as the measure predicting the incremental response to an action. It is the difference in decision probability with and without treatment. Instances are categorized into control and exposed groups, with the former exposed to the recommended action and the latter not.

%% file: Tikz/illustration.tex
\begin{tikzpicture}
    \tikzstyle{triangle}=[star, star points=5, draw=black, star point ratio=0.55, 
    fill=red, shape border rotate=180];
    \def\p{figures/toy};
    
    \node at (0, 0) {\includegraphics[width=8cm]{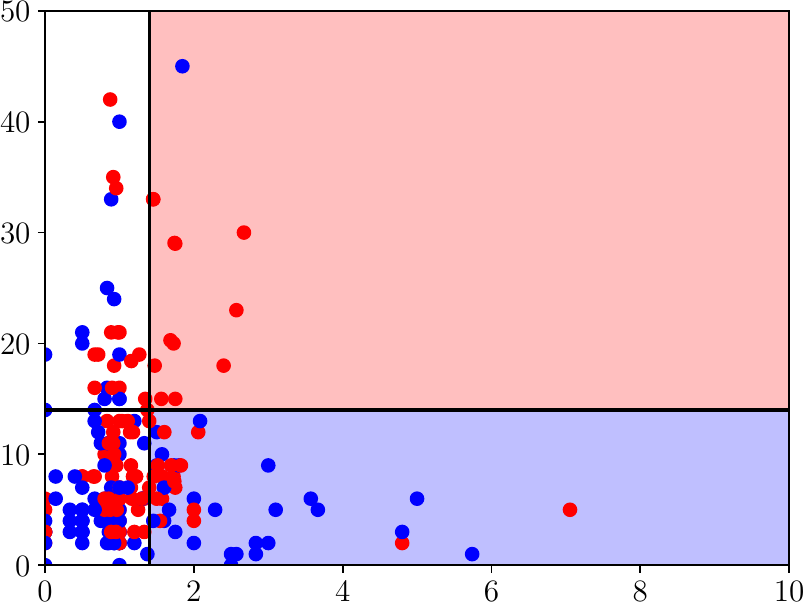}};
    \node at (0.2+0,-3.25) {avg\_cc};
    \node[rotate=90] at (-4.5+0,0.2) {cbo};
    
    \node[align=left, scale=0.85] at (0+1.1,1) {
    $[(\text{avg\_cc}>1.4) \land (\text{cbo}>14)] \Rightarrow \text{bug}$\\
    Support 6\% Confidence 91\%};
    \node[align=left, scale=0.85] at (0+1.3,-1.45) {
    $[(\text{avg\_cc}>1.4) \land (\text{cbo}\leq 14)] \Rightarrow \text{no-bug}$\\
    Support 13\% Confidence 58\%};
    \node[triangle, scale=0.75] at (-2+0,-0.5) {File};
\end{tikzpicture}

%% file: sections/2_methodology.tex
\section{Methodology}
\label{sec:setup}

\textbf{Overview.} We introduce a novel defect reduction planning approach named CounterACT, based on Action Rule Mining (ARM) leveraging Counterfactuals. The ARM algorithm ~\cite{Sykora2020ActionRC} is designed to mine rules on nominal data, hence our approach adds data discretization using entropy \cite{Fayyad1993MultiIntervalDO} to transform numerical values into tuples which enables the processing of numerical data, expanding the applicability of the approach. In addition, we add actionable analysis detailed in \ref{subsec:actionable_analysis} for the selection of actionable features. ARM generates multiple defect reduction plans when given a defective instance. Thus, we augment the algorithm with a plan selection algorithm presented in Section \ref{subsec:plan_selection} that leverages historical data based on overlap to ensure precedence and improve the plausibility of plans. 
\begin{figure}[!h]
\vspace{-15mm}
\centering
\centerline{\includegraphics[width=0.95\textwidth]{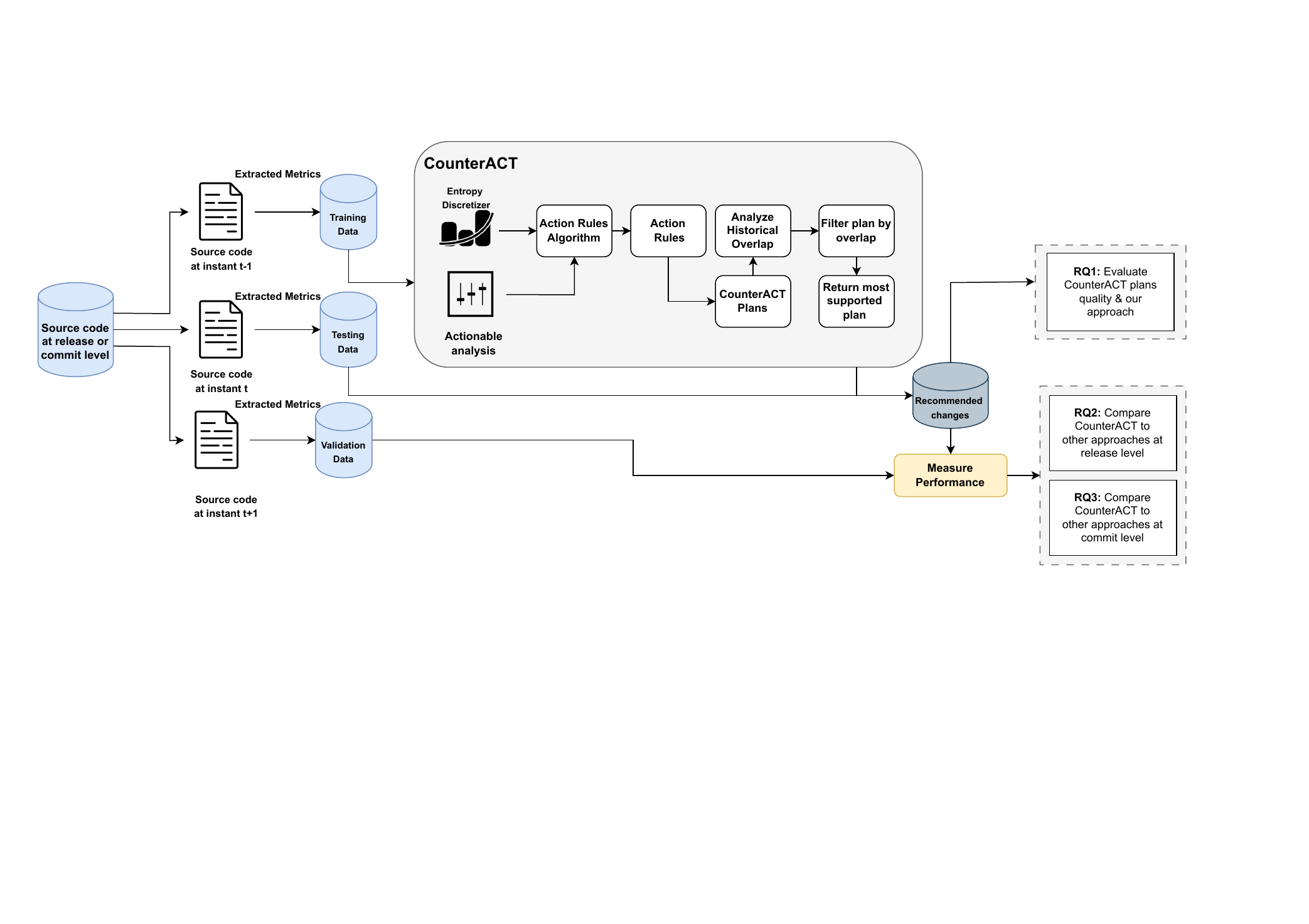}}
\vspace{-39 mm}

\caption{\texttt{CounterACT}: Overview of the approach. Note that for evaluating other benchmark planners, the grey area be replaced by the corresponding planner.}
\label{fig:overview_counterACT}
\vspace{-5mm}
\end{figure}
We benchmarked the
performance of CounterACT on the release level against the original ARM algorithm and several state-of-the-art approaches: model-agnostic techniques
such as TimeLime~\cite{peng2021defect} and Lime \cite{Jiarpakdee}, XTree~\cite{krishna2020prediction}, and statistical methods including
Alves, Shatnawi, and Oliveira~\cite{alves2010deriving,shatnawi2010quantitative,oliveira2014extracting}. In addition, we ventured a step
further, examining CounterACT’s planning at the commit level and measuring its performance with regard to the aforementioned approaches.

\begin{figure}[!h]
\centering
\centerline{\includegraphics[width=0.8\textwidth]{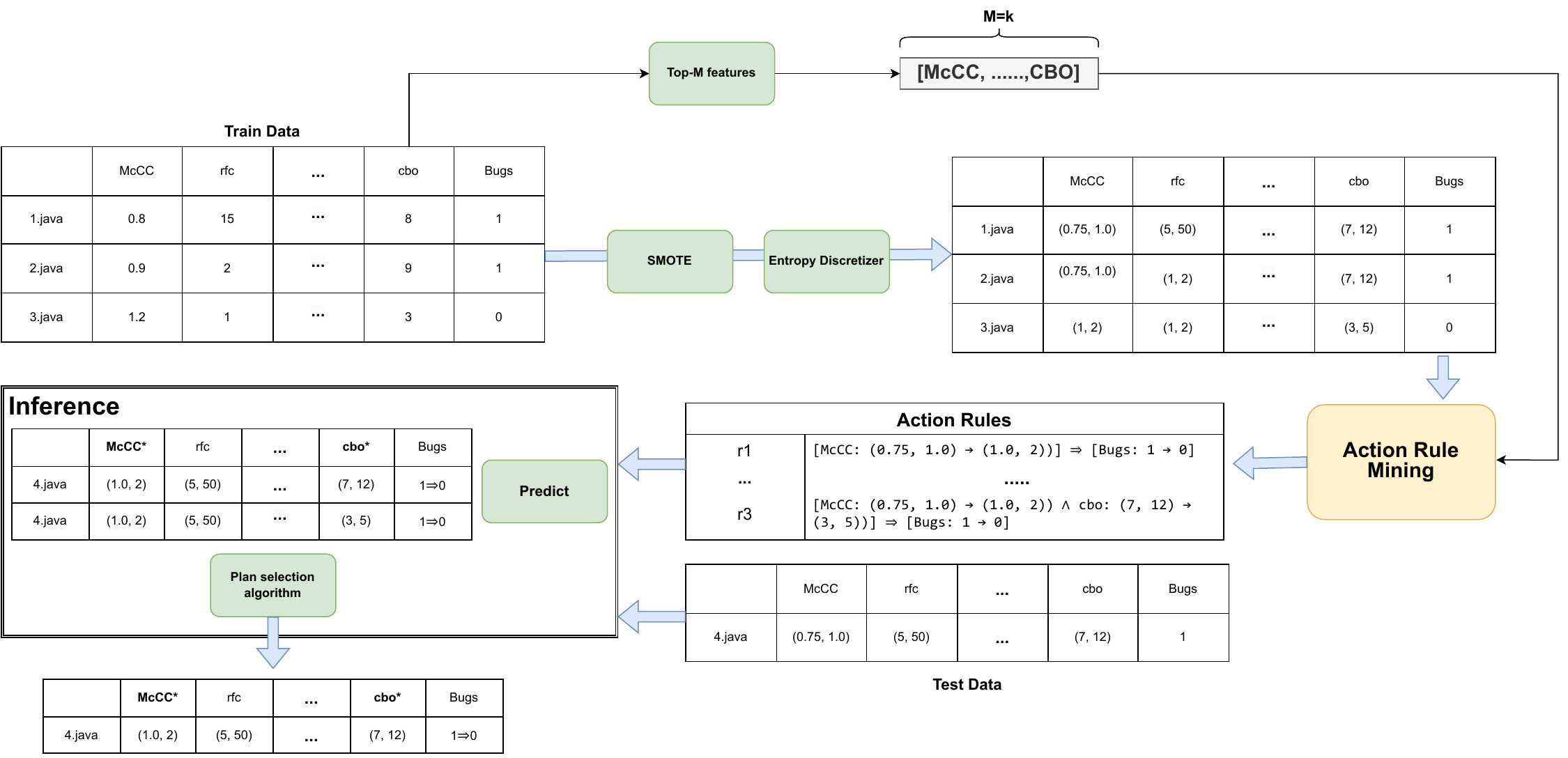}}
\vspace{-2mm}
\caption{CounterACT: an illustrative example}

\label{fig:counterACT_train}
\vspace{-5mm}
\end{figure}

Figure \ref{fig:overview_counterACT} and \ref{fig:counterACT_train} illustrate an overview of the CounterACT approach. It consists of the following steps:
(1) Actionable Analysis; (2) Mining action rules from historical data; : (3) plan selection.

\subsection{Actionable Analysis} 
\label{subsec:actionable_analysis}

Building upon the foundations set by previous research~\cite{7426628, peng2021defect}, which mainly restricts the generated 
recommendations to the attributes that were seen to be frequently modified within the history of a software 
project. We operate under the assumption that any suggested alterations to a code metric should align with 
observations from historical records.

Consequently, we generate the actionable set encompassing the top-M features with the highest variance in historical data. Here, the user has the flexibility to specify the value of M. 
In the context of our experiments, the selection of the top-M features is guided by evaluating the magnitude of changes observed between two consecutive releases ($t-1$, $t$). This is determined by measuring the hedge $g$ value~\cite{Kampenes}, which is calculated as:
\begin{equation}
    g = \big(\mu^{[t-1]} - \mu^{[t]}\big)/{S_{\text{pooled}}},
    \label{eq:hedge}
\end{equation}
and
\begin{equation}
    S_{\text{pooled}} = \sqrt{
    \frac{(n^{[t-1]} - 1)(\sigma^{[t-1]})^2 + (n^{[t]} - 1)(\sigma^{[t]})^2}
    {n^{[t-1]} + n^{[t]} - 2}},
    \label{eq:pooled}
\end{equation}
Here, the variables $\mu^{[t-1]},\mu^{[t]}$, $\sigma^{[t-1]},\sigma^{[t]}$ represent respectively the means and standard deviations of an attribute in two
consecutive releases. The terms $n^{[t-1]}$ and $n^{[t]}$ denote the sample size for each release. Moreover, since a plan is generated using the set of top-M features with the most changes, we denote the latter as the actionable set. This means that the action rule algorithm will mine rules altering only the features in that set. 

\subsection{Mining Action Rules}

\textbf{Motivation. } Among the variety of recommendation techniques developed by the ML community, 
Action Rules \cite{Sykora2020ActionRC} are most relevant to our problem for multiple reasons. 
First, most Counterfactual methods return a single perturbed instance $x'$ 
\cite{wachter2017counterfactual,parmentier2021optimal} 
or a set of $k$ perturbed instances 
$\{x'_i\}_{i=1}^k$\cite{russell2019efficient,mothilal2020explaining}. 
However, in the context of software engineering, we want recommendations from $x$ to $x'$ to match 
past changes in code. If the code metrics stored in the vector $x$ are numerical, then it is not realistic 
to expect to perceive a change that goes exactly from $x$ to $x'$ in the code history.
To ensure that overlaps with previous code changes are non-zero, like previous defect reduction 
planners proposed in software engineering, it is better to provide a recommended \emph{range} of feature values. 
For example, instead of recommending changing a code metric $x$ from $x=34$ to $x=25$, we want to 
recommend reducing $x=34$ to a range $10\leq x\leq 25$, which is exactly what action rules provide. 
Secondly, the plans generated by action rule miners are transparent.
Indeed, since these action rules are derived from classification rule mining, their computation 
is well documented \cite{10.1007/3-540-45329-6_8}, and, assuming they are small enough, the rules 
could be visualized and understood by software developers. 

Since the action rule mining algorithm takes discretized ranges of attribute values, drawing 
inspiration from previous studies such as TimeLIME~\cite{peng2021defect} and LIME~\cite{ribeiro2016should}, we use an entropy-based discretizer \cite{Fayyad1993MultiIntervalDO} 
which provides a discretized interval indicating the range of values during which the feature will 
maintain the same effect to the prediction result.

After obtaining the set of actionable features, we set them as flexible attributes in the algorithm of 
action rule mining. Then, we train the algorithm on historical releases 
(RQ2) or commits (RQ3). We use the original algorithm's default values for the minimum support (5\%) and minimum confidence 
(55\%). The latter refers to those of the classification rules from which action rules are generated.

\textbf{Quality assessment of action rules. }In addition to the evaluation of the generated action 
rules in terms of support, confidence, and uplift, we further extend the assessment by calculating the following metrics:
\begin{itemize}
    \item \textbf{False Positive Action Rule (FP):} An action rule is considered \textbf{\texttt{FP}} if the suggested action for the metrics has been historically observed not to fix a bug. In other words, if the historical change between versions $t-1$ and $t$ does not resolve a bug in a file and the action rule matches this change, then the rule is considered as a \textbf{\texttt{FP}}.
    \item \textbf{True Positive Action Rule (TP):} if the suggested action for the metrics has been historically observed to fix a bug. In other words, if the historical change between versions $t-1$ and $t$ did resolve a bug in a file and the action rule matches this change, then the rule is considered as a \textbf{\texttt{TP}}.
\end{itemize}

\subsection{Plan Selection:}
\label{subsec:plan_selection}

The last step of the action rule mining algorithm is inference. In which we input a defective instance and the algorithm returns a recommended action on each software attribute inferred from the previously mined action rules. We present two examples of defect reduction plans $(p_1)$, $(p_2)$:

\begin{equation}
    \begin{aligned}
    \text{(Action Rule)\,\,\,\,} &p1\,\, = [NOC\_class (0, 0.5) \land NUMPAR\_method((3.0, 4.0) \to  (1.0, 3.0)), \\
    &NL\_method( (0.5, 1.0) \to (0.0, 0.5) ), \\    
    &McCC\_method( (0.99, 1.0) \to   (0.0, 0.99))] \Rightarrow [True \to False]\\
    \text{(Action Rule)\,\,\,\,} &p2\,\, = [NOC\_class(0.5, 1.0) \land NOI\_method((0.0, 0.5) \to  (0.0, 1.0)), \\
    & NUMPAR\_method((3.0, 4.0) \to  (1.0, 3.0)) \Rightarrow [True \to False]    
    \end{aligned}
\label{eq:action_rule_def}
\end{equation}

\input{sections/plan_select_algorithm}

Since we have multiple plans, we establish a plan selection process that leverages the similarity to historical records by aligning the selected plans more closely with the proven historical patterns of defect bug reductions.
Initially, a set of candidate plans is generated. Then we measure the overlap of each plan with the historical logs using 
Equation \ref{eq:overlap} and return the plan with the highest overlap score.

\begin{figure}[!h]
\vspace{-3mm}
    \centering
    \begin{minipage}{0.5\textwidth}
        \centering
        \includegraphics[width=0.95\linewidth]{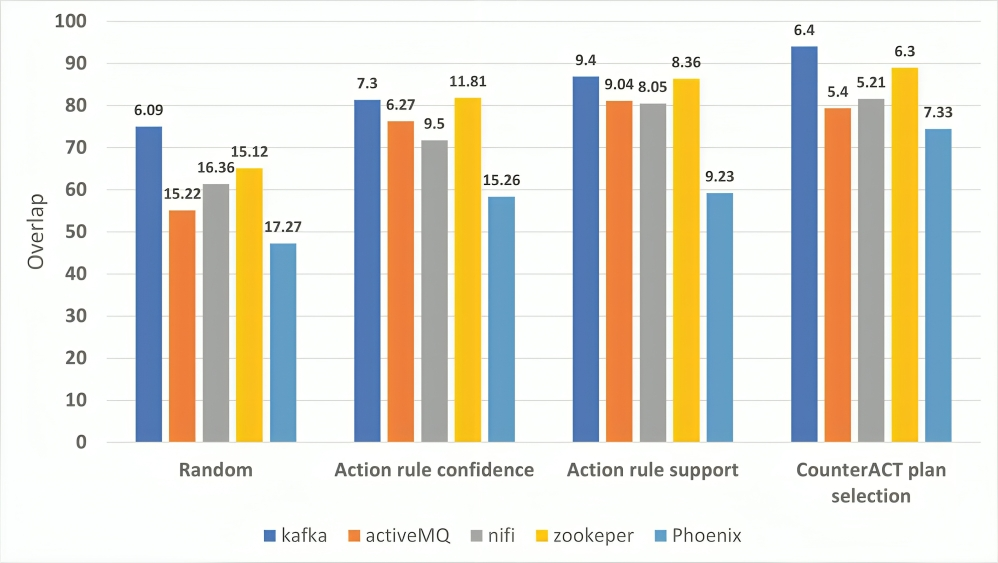}
        \caption{ Median overlap scores between different \\ plan selection methods. The reported values \\ on the histograms are the standard deviations}
        \label{figure:plan_selec}
    \end{minipage}%
    \begin{minipage}{0.5\textwidth}
        \centering
        \includegraphics[width=\linewidth]{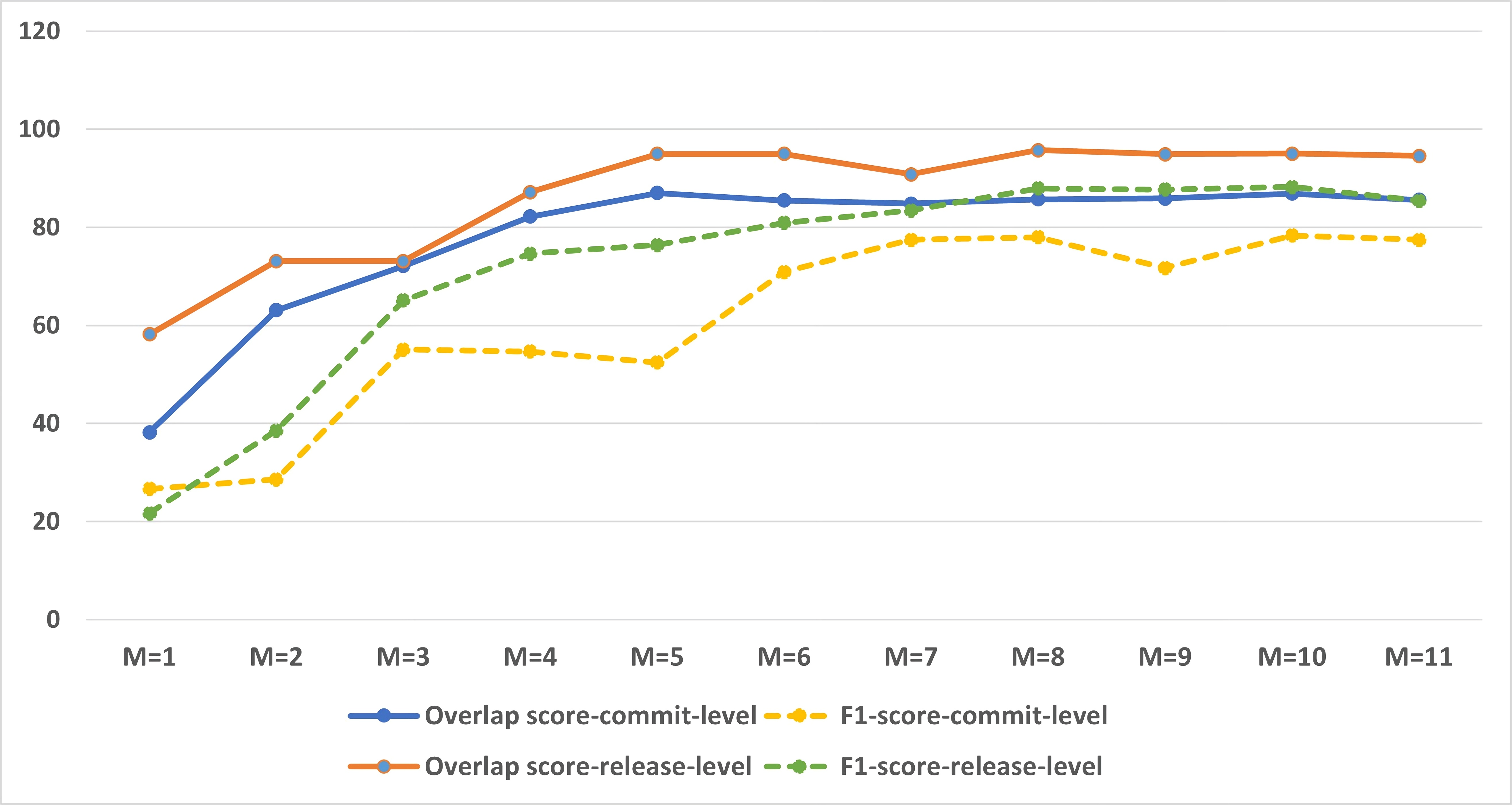}
        \caption{Results of top-M experiments. We chose the value of $M=10$, since it shows the most stable overlap and F1 score}
        \label{figure:M}
    \end{minipage}
    \vspace{-6mm}

\end{figure}

Moreover, we perform an analysis to assess the impact of different plan selection strategies for the buggy code instances. Our approach generates multiple defect reduction plans, thus we evaluate the outcomes of selecting a plan:  (a) at random, (b) based on the highest action rule support, (c) based on the highest action rule confidence, and (d) our proposed plan selection. In Figure \ref{figure:plan_selec}, we evaluate the different plan selection approaches and report the median overlap score. On average, our plan-selection approach results in the highest median overlap score of 86.14\%. In comparison, relying on the action rule support results in a median overlap score of 85.5\%.

Figure \ref{figure:plan_selec} shows that our plan selection algorithm performs better overall. In addition, after conducting Wilcoxon signed rank statistical test \cite{woolson2007wilcoxon} to compare the values of the various strategies against the ’CounterACT plan selection’, the results indicate that ’Random’ and the ’action rule confidence’ values both show statistically significant differences from the ’CounterACT plan selection’, with p-values less than 0.05. However, the ’action rule support’ values do not show a statistically significant difference from ’CounterACT plan selection’, as evidenced by a p-value of approximately 0.3123. While the overlap of CounterACT’s plan selection is not significantly higher than the ’action rule support’ we still favor it, because it does generally provide better overlap score and more stable results as evidenced by the lower standard deviation. For a comprehensive breakdown of the underlying algorithm to select the most supported plan, please refer to 
Algorithm~\ref{alg:palnSelection}.

\section{Experimental details}
\label{subsec:design}

The experiments report the performance of CounterACT and
other state-of-the-art approaches by comparing the quality of
plans recommended by each approach at the release and commit levels.

\subsection{Release Level}

\textbf{Studied projects:} To conduct our study, we use a publicly available defect dataset used in previous work \cite{peng2021defect,krishna2020prediction} to support verifiability and foster replicability. The data consists of 9 Apache open-source projects collected by Jureczko et al~\cite{10.1145/1868328.1868342}
as described in Table \ref{tab:project_desc} (Camel was used twice since we have 4 releases). It is necessary for the data to contain a minimum of three consecutive releases for each project since the main hypothesis of our approach is that the guidance which is derived from past knowledge (a release t-1) can be used to provide plans for defective files in
the target releases (a release $t$) and be applicable to
prevent software defects in future releases (a release
$t + 1$).  

\input{tables/project_desc}

\input{tables/cko}

\noindent
\textbf{Software metrics: }The data consists of records of previous defects for software modules (e.g. classes) for which there are 20 static code metrics that quantify the aspects of software
design as shown in Table \ref{tab:cko}. They are a combination of McCabe \cite{1702388} and complexity metrics in conjunction with extended CKOO metrics \cite{295895} and Lines of Code (LOC). Previous studies indicate that these measures correlate highly with the likelihood of defects and provide a comprehensive analysis of the code complexity and object-oriented aspects \cite{article} where they analyzed the significance and correlation of each software metric with defect numbers.

\textbf{Actionable analysis: }
In order to get the set of actionable attributes, we input historical data from the older releases to compute the variance of each feature. Then we select the top-M features with the largest variance as the actionable set of flexible attributes. Recommendations based on other features are ignored. The parameter M can be user-specified and the features may vary with respect to different projects and the releases used as historical data. Here we set the default value of M to be 10, which means 50\% of all twenty features can be mutated.

We experiment with various values of M to determine the one that yields the best overlap score as an evaluation metric on the release and commit level data, ensuring an optimal setting is found through exhaustive evaluation. As shown in Figure \ref{figure:M}, our results suggest that M = 10 is a useful default setting as it provides the more stable median F1-score (88\%) and highest overlap (95\%) for the release level data, and yields a stable median F1-score and overlap score with respective values of 78\% and 85.97\% for the commit level data.


\textbf{Mining action rules: }
First, we use an over-sampling technique named SMOTE (Synthetic Minority Oversampling Technique) \cite{smote} since we work on imbalanced datasets where defective instances constitute only a minor proportion of the entire data set. As suggested by prior work \cite{DBLP:journals/corr/AgrawalM17, 8494821} who found that the SMOTE technique outperforms other class rebalancing techniques, SMOTE finds its nearest neighbors in the feature space for each instance in the minority class. Then, it randomly selects some of these nearest neighbors and generates synthetic instances. It does this by interpolating between a minority class instance and its selected neighbor. Finally, SMOTE combines the synthetic instances with the undersampling of the majority class to produce a set of balanced instances. We use the implementation of SMOTE as provided by Imbalanced-Learn Python library \cite{Lematre2016ImbalancedlearnAP}. This step is crucial because previous research on the studied dataset has highlighted the difficulty of building predictors when dealing with small target classes \cite{DBLP:journals/corr/AgrawalM17}.

Next, we use an entropy-based discretizer on the data and train the action rule mining algorithm on release t-1 and release t. 
Then we conduct inference, where we generate defect reduction plans on the defective instances of release t and then compute the overlap with actual changes made by developers in the future release t+1. 

After getting defect reduction plans from the planners, we assess their performance
using the overlap score as described in Section \ref{sec:background}. In addition, to assess whether a plan has a high/low similarity to a bug reducing/adding actions. We measure the improvement score, also detailed in Section \ref{sec:background}.  Given that the total number of bugs varies from each project as shown in Table \ref{tab:project_desc}, we compare the results proportionally to their respective project. Indeed, we expect a project with more
bugs reduced in the validation dataset (i.e release t+1) to result in a planner that scores higher than a dataset
with fewer bugs reduced. 
\subsection{Commit-level}
In this section, we detail the experimental setup for examining CounterACT’s planning at the commit level, leveraging counterfactual-based planning at a finer-grained level (change-level).

\textbf{Data:}
 We select just-in-time defect datasets from the 2021 PROMISE file-level just-in-time defect prediction challenge \cite{Trautsch_PROMISE_2021_Defect_2021}. Thus, our data contains one instance for every file that was changed within each commit. We choose to work with this data since it consists of different large-scale open-source Apache projects and the labels were assigned based on specific criteria: commits were tagged as bug-fixing if they included a link to a Jira issue, the issue was identified as a bug, and the fastText classification method \cite{DBLP:journals/corr/abs-2003-05357}, with a 95\% recall rate, verified this. The authors of the data used git blames to determine the inducing changes for each changed line in a Java file using the SZZ algorithm \cite{inproceedings}. To avoid selection bias, We randomly select 5 projects, as detailed in Table \ref{tab:project_desc_commit}.
 
\input{tables/project_desc_commit}

\textbf{Commit features: } The dataset span across 5 categories of many fine-grained features: Just-In-Time (JIT)  \cite{6341763}, Fine-Grained JIT \cite{PASCARELLA201922}, PMD features, Warning density features, and static code metrics. We focus on static code metrics since the criteria we use are that (1): static code metrics are widely used in defect prediction \cite{DBLP:journals/corr/abs-2109-03544}, (2): in our study, we do not only look to provide to explain "why" an instance was defective but we also provide "what to do". For example, while JIT features are widely used for commit-level defect prediction studies, static code metrics provide a descriptive analysis of the code and how complex its structure gets. We adopt code metrics that cover the source code properties of complexity, coupling, documentation, inheritance, and size as shown in Table \ref{tab:cko}. Moreover, we mitigate collinearity and multi-collinearity by using AutoSpearman \cite{autospearman},
an automated feature selection approach. After using the latter, we finally select 18 features that are not highly correlated with each other. To handle the class imbalance we apply SMOTE, only
on the training dataset. 

\textbf{Actionable analysis: }
Following the same approach as our release level analysis, we set the number of flexible attributes to the action rule mining algorithm using the top M=10 features with the highest variance between windows of commits. We first sort the commits by date. Subsequently, we partition the data into three equal chunks. We then compute the variance for each feature between these chunks. Note that we set the number of changed features of the other planners the same as CounterACT planner to ensure fair comparison for both release and commit level experiments.

\textbf{Mining action rules} In our study, we proceed as follows to split our data into train, test, and validation sets:
\begin{itemize}
    \item Training Set: we start by chronologically ordering all commits. We excluded commits from the most recent three-month window to avoid the inclusion of recent buggy commits that may lack associated fixes. Following this, we extracted the last 250 commits (in alignment with the criteria proposed by the PROMISE challenge) to create our test set. All the historical commits represent our training set.
    \item Test Set: This split is composed exclusively of commits identified as "buggy" from the reserved 250 commits.
    \item Validation Set: For each "buggy" commit within the test set, we identify a corresponding "fix" commit as follows: The "fix" commit should relate to the same file as its associated buggy commit. The "fix" commit must have a commit date succeeding that of its corresponding buggy commit. Finally, the identified `"fix" commit must be marked as non-buggy.
\end{itemize}

%% file: sections/plan_select_algorithm.tex
\RestyleAlgo{ruled}
\SetKwComment{Comment}{ }{ */}

\begin{algorithm}[htb]
    \caption{Plan Selection Algorithm}
\KwIn{Candidate plans $generated\_plans$ and non-buggy historical data $past\_fixes$}
\KwOut{The proposed $plan$ for the instance}
$plan \gets None$\;
$pool\_overlap\_score\_list \gets list()$\;
\For{$plan\_candidate$ in $generated\_plans$}{
    $overlap\_score\_list \gets list()$\;    
    \For{$fix$ in $past\_fixes$}{
        $overlap\_score  \gets Overlap(plan\_candidate, fix)$\;        
        append $overlap\_score$ into $overlap\_score\_list$\;        
    }
    $plan\_overlap$  $\gets$ median($overlap\_score\_list$)\;  
    append $plan\_overlap$ into $pool\_overlap\_score\_list$\;
}
$max\_overlap\_index$ $\gets$ random index of max value in $pool\_overlap\_score\_list$\;
$plan$ $\gets$ element at $max\_overlap\_index$ from $generated\_plans$\;

\label{alg:palnSelection}
\vspace{-1mm}
\end{algorithm}

%% file: tables/project_desc.tex

\begin{table}[!ht]
\vspace{-2mm}
\caption{The dataset of defects used. The last column displays the decrease in bugs found in identical files when comparing the test release to the validation release. A negative number in this row suggests that the subsequent validation release had a higher bug count than its predecessor.}
\vspace{-2mm}
\scalebox{0.8}{
\begin{tabular}{lcccccccc}
\hline
\textbf{Project} &
  \multicolumn{1}{l}{\textbf{\begin{tabular}[c]{@{}l@{}}Training \\ (oldest)\end{tabular}}} &
  \multicolumn{1}{l}{\textbf{\begin{tabular}[c]{@{}l@{}}Testing \\ (newest)\end{tabular}}} &
  \multicolumn{1}{l}{\textbf{\begin{tabular}[c]{@{}l@{}}Validation\\ (most recent)\end{tabular}}} &
  \multicolumn{1}{l}{\textbf{\#files}} &
  \multicolumn{1}{l}{\textbf{\begin{tabular}[c]{@{}l@{}}\# matched \\ files\end{tabular}}} &
  \multicolumn{1}{l}{\textbf{\begin{tabular}[c]{@{}l@{}}\# bugs in \\ testing set\end{tabular}}} &
  \multicolumn{1}{l}{\textbf{\begin{tabular}[c]{@{}l@{}}\# bug in \\ validation set\end{tabular}}} &
  \multicolumn{1}{l}{\textbf{\begin{tabular}[c]{@{}l@{}}\# bugs \\ reduced\end{tabular}}} \\ \hline
Jedit    & 4.0 & 4.1 & 4.2 & 367 & 78  & 216 & 74  & 142 \\ \hline
Camel    & 1.0 & 1.2 & 1.4 & 872 & 210 & 508 & 247 & 261 \\ \hline
log4j    & 1.0 & 1.1 & 1.2 & 205 & 35  & 83  & 120 & -37 \\ \hline
Xalan    & 2.5 & 2.6 & 2.7 & 885 & 385 & 529 & 381 & 148 \\ \hline
Ant      & 1.5 & 1.6 & 1.7 & 745 & 91  & 183 & 163 & 20  \\ \hline
Velocity & 1.4 & 1.5 & 1.6 & 229 & 138 & 321 & 144 & 177 \\ \hline
Poi      & 1.5 & 2.5 & 3.0 & 442 & 247 & 495 & 366 & 129 \\ \hline
Synapse  & 1.0 & 1.1 & 1.2 & 256 & 58  & 97  & 65  & 32  \\ \hline
\end{tabular}
}
 \vspace{-7mm}
\label{tab:project_desc}
\end{table}

%% file: tables/cko.tex
\begin{table}[!h]
\caption{Software metrics.}
\vspace{-4mm}

\scalebox{0.69}{
\begin{tabular}{ll|lll}
\hline
\multicolumn{2}{c|}{\textbf{Release level}} & \multicolumn{3}{c}{\textbf{Commit level}} \\ \hline
\multicolumn{1}{l|}{\textbf{Metric}} & \textbf{Description} & \multicolumn{1}{l|}{\textbf{Granularity}} & \multicolumn{1}{l|}{\textbf{Metric}} & \textbf{Description} \\ \hline
\multicolumn{1}{l|}{WMC} & Number of methods & \multicolumn{1}{l|}{\multirow{11}{*}{\textbf{Class}}} & \multicolumn{1}{l|}{CBO} & Coupling Between Object classes \\
\multicolumn{1}{l|}{DIT} & Depth of Inheritance Tree & \multicolumn{1}{l|}{} & \multicolumn{1}{l|}{TNG} & Total Number of Getters \\
\multicolumn{1}{l|}{NOC} & Number of Children & \multicolumn{1}{l|}{} & \multicolumn{1}{l|}{NLPA} & Number of Local Public Attributes \\
\multicolumn{1}{l|}{CBO} & Number of objects the class is coupled with & \multicolumn{1}{l|}{} & \multicolumn{1}{l|}{NLE} & Nesting Level else if \\
\multicolumn{1}{l|}{RFC} & Response for a Class & \multicolumn{1}{l|}{} & \multicolumn{1}{l|}{NOD} & Number of Descendants \\
\multicolumn{1}{l|}{LCOM} & \begin{tabular}[c]{@{}l@{}}Lack of cohesion in methods\end{tabular} & \multicolumn{1}{l|}{} & \multicolumn{1}{l|}{NII} & Number of Incoming Invocations \\
\multicolumn{1}{l|}{CA} & Number of Afferent Couplings & \multicolumn{1}{l|}{} & \multicolumn{1}{l|}{TNLS} & Total Number of Local Setters \\
\multicolumn{1}{l|}{CE} & Number of Efferent Couplings & \multicolumn{1}{l|}{} & \multicolumn{1}{l|}{NOP} & Number of Parents \\
\multicolumn{1}{l|}{LCOM3} & Normalized measure of Lack of Cohesion & \multicolumn{1}{l|}{} & \multicolumn{1}{l|}{LCOM5} & Lack of Cohesion in Methods 5 \\
\multicolumn{1}{l|}{LOC} & Size of the Java Byte Code & \multicolumn{1}{l|}{} & \multicolumn{1}{l|}{CI} & Clone Instances \\
\multicolumn{1}{l|}{DAM} & Data access & \multicolumn{1}{l|}{} & \multicolumn{1}{l|}{NUMPAR} & Number of Parameters \\ \cline{3-5} 
\multicolumn{1}{l|}{MOA} & Number of attributes which are user-defined classes & \multicolumn{1}{l|}{\multirow{9}{*}{\textbf{Method}}} & \multicolumn{1}{l|}{CI} & Clone Instances \\
\multicolumn{1}{l|}{MFA} & Ratio of method which are inherited & \multicolumn{1}{l|}{} & \multicolumn{1}{l|}{McCC} & McCabe's Cyclomatic Complexity \\
\multicolumn{1}{l|}{NPM} & Number of Public Method (API Size) & \multicolumn{1}{l|}{} & \multicolumn{1}{l|}{NOI} & Number of Outgoing Invocations \\
\multicolumn{1}{l|}{CAM} & Cohesion Amongst Classes & \multicolumn{1}{l|}{} & \multicolumn{1}{l|}{NLE} & Nesting Level else if \\
\multicolumn{1}{l|}{IC} & Inheritance coupling & \multicolumn{1}{l|}{} & \multicolumn{1}{l|}{NII} & Number of Incoming Invocations \\
\multicolumn{1}{l|}{CBM} & Coupling between methods & \multicolumn{1}{l|}{} & \multicolumn{1}{l|}{LLOC} & Logical Lines of code \\
\multicolumn{1}{l|}{AMC} & Average method complexity & \multicolumn{1}{l|}{} & \multicolumn{1}{l|}{TNS} & Total Number of Statements \\
\multicolumn{1}{l|}{max\_CC} & Maximum McCabe’s cyclomatic complexity seen in class & \multicolumn{1}{l|}{} & \multicolumn{1}{l|}{} &  \\
\multicolumn{1}{l|}{avg\_CC} & Average McCabe’s cyclomatic complexity seen in class & \multicolumn{1}{l|}{} & \multicolumn{1}{l|}{} &  \\ \hline
\end{tabular}
}
\label{tab:cko}
\vspace{-3mm}
\end{table}

%% file: tables/project_desc_commit.tex
\begin{table}[!h]
\vspace{-2mm}
\caption{A  summary of the studied commit defective datasets provided by PROMISE challenge.}
\vspace{-2mm}
\scalebox{0.63}{
    \begin{tabular}{l|cccc|cccc}
    \hline
    \multirow{2}{*}{\textbf{Projects}} & \multicolumn{4}{c|}{\textbf{Training Data}}     & \multicolumn{4}{c}{\textbf{Testing Data}}     \\ \cline{2-9} 
     &
      \multicolumn{1}{l}{\textbf{Start Date}} &
      \multicolumn{1}{l}{\textbf{End Date}} &
      \multicolumn{1}{l}{\textbf{\# Commits}} &
      \multicolumn{1}{l|}{\textbf{\# Defective Commits}} &
      \multicolumn{1}{l}{\textbf{Start Date}} &
      \multicolumn{1}{l}{\textbf{End Date}} &
      \multicolumn{1}{l}{\textbf{\# Commits}} &
      \multicolumn{1}{l}{\textbf{\# Defective Commits}} \\ \hline
    nifi                               & 08/12/2014 & 17/02/2017 & 16445 & 1557 (9.4\%)  & 17/05/2017 & 29/09/2017 & 1093 & 145 (13.2\%) \\ \hline
    zookeper                           & 03/11/2007 & 01/07/2013 & 3545  & 422 (11.9\%)  & 07/10/2013 & 11/09/2017 & 850  & 89 (10.4\%)  \\ \hline
    kafka                              & 01/08/2011 & 15/03/2017 & 6742  & 818 (12.1\%)  & 16/06/2017 & 29/09/2017 & 1244 & 99 (7.9\%)   \\ \hline
    Phoenix                            & 27/012014  & 17/08/2016 & 8833  & 1486 (16.8\%) & 17/11/2016 & 28/09/2017 & 1094 & 196 (17.9\%) \\ \hline
    ActiveMQ                           & 12/12/2005 & 11/04/2016 & 28570 & 3568 (12.4\%) & 11/07/2016 & 21/09/2017 & 1051 & 73 (6.9\%)   \\ \hline
    \end{tabular}
}
\label{tab:project_desc_commit}
\vspace{-4mm}
\end{table}

%% file: sections/3_results.tex
\section{Results}
\label{sec:results}

\input{sections/RQ1}
\input{sections/RQ2}

\input{sections/RQ3}

%% file: sections/RQ1.tex
\vspace{-3pt}
\subsection*{\textbf{RQ1: How effective is the rule based guidance generated by CounterACT?}}
\setcounter{subsubsection}{0}



\begin{figure}[!h]
    \vspace{-2mm}
    \centering
    \begin{minipage}{0.5\textwidth}
        \centering
        \includegraphics[width=0.75\linewidth]{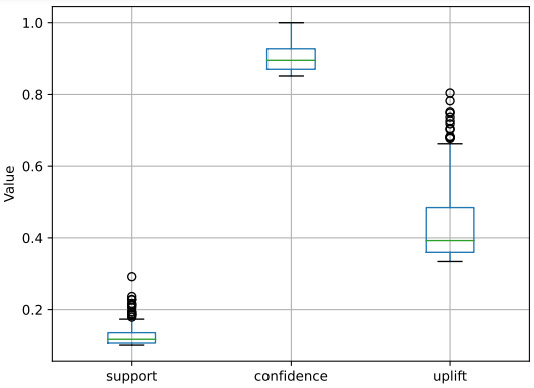}
        \vspace{-3mm}
        \caption{Support, Confidence, Uplift distribution \\at the release level}
        \label{figure:supp_conf}
    \end{minipage}%
    \begin{minipage}{0.5\textwidth}
        \centering
        \includegraphics[width=0.75\linewidth]{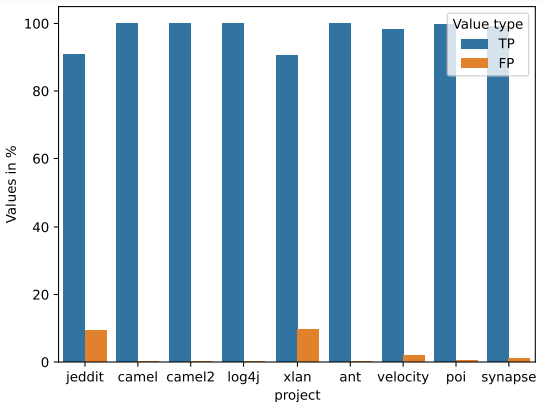}
        \vspace{-3mm}
        \caption{The True Positive (TP) and False Positive (FP) of the learned rules}
        \label{figure:tp_fp}
    \end{minipage}
    \vspace{-3mm}
\end{figure}

\indent
In figure~\ref{figure:supp_conf} we illustrate the distribution of support, confidence and uplift of the action rules generated on the release level of all projects. CounterACT produces rules characterized by high confidence (average 89\%), indicating a high conditional probability that when a bug-reducing action is recommended by the action rule, the predicted non-buggy outcome is likely to occur in 89\% of cases. However, average support shows that approximately 13\% of the observed bugs have a specific associated recommended fix. Given that the support values range from 10.12\% to 29.18\%, this suggests that some fixes are applicable to a wide range of bugs, while others might be more specific or less common. From this, we can infer that many rules, while specific and relevant to a minor group of bugs, remain trustworthy because of the high confidence under specific conditions. Moreover, the rules tend to have a high uplift value of 43\% and we observe that all the values are positive. Following our definition in section~\ref{sec:actionrulemining}, the uplift indicates that the occurrence of the action to fix a bug (treatment) increases the chances of the bug fix happening more than if we did not apply any action. This means that the rules are not only capturing frequent bug fix actions but also actions that have a meaningful relationship with the outcome.

\textbf{The majority of learned action rules are associated with bug fixes.}
In Figure \ref{figure:tp_fp}, we display the True Positive (TP) and False Positive (FP) rates for the learned rules across all projects. We observe that CounterACT achieves a TP rate higher than 80\%  for each project. Indeed, this implies that the action rules mined by CounterACT are often present in the history of actual bug reduction plans used by developers.

\vspace{-2mm}
\begin{tcolorbox}
\vspace{-2mm}
CounterACT produce rules that, while based on less frequent patterns (13\%), are highly predictive (high confidence of 89\%) and indicate a meaningful relationship between antecedent and consequent (significant uplift of 43\%). Additionally, the learned action rules are highly associated with historical bug fixes done by developers (average TP 97.56\%).
\vspace{-2mm}
\end{tcolorbox}

%% file: sections/RQ2.tex
\vspace{-3pt}
\subsection*{\textbf{RQ2: How does CounterACT compare against competing defect reduction approaches at the release level?}}
\label{sec:RQ2}
\setcounter{subsubsection}{0}



\indent

The recommendations provided by CounterACT align more closely with developers' actions compared to competing approaches. 
This is evidenced by a higher overlap score : 95\% on average. Tables \ref{tab:overlap} and \ref{tab;IQR} demonstrate how 
frequently code changes intersect with CounterACT's suggested plans within projects. The median overlap score, 
$O(p,m^{[t]},m^{[t+1]})$, for each planner across all instances are reported in Table \ref{tab:overlap}. In Table \ref{tab;IQR}, we present the interquartile range (IQR) for overlap scores among all plans. When median overlap scores are similar, a lower IQR signifies greater stability and robustness in the defect reduction planner. CounterACT shows a small IQR (5 on average) while maintaining the highest median overlap score in 8 out of 9 projects when compared to TimeLIME, the current state-of-the-art. This shows that using action rules generates robust plans and prevails over other defect reduction planners in terms of providing plans that better resemble developers’ choices, all without relying on a black box model. Thus, CounterACT is designed to provide achievable and maintainable solutions, which is not the case for ARM algorithm. As shown in Table \ref{tab:overlap}, the ARM algorithm on its own was unable to generate rules for 5 out of 9 projects on the release level, having an median overlap score of 31.55\%.


\input{tables/overlap}

\begin{figure}[!h]
    \centering
    \centerline{\includegraphics[width=1\textwidth]{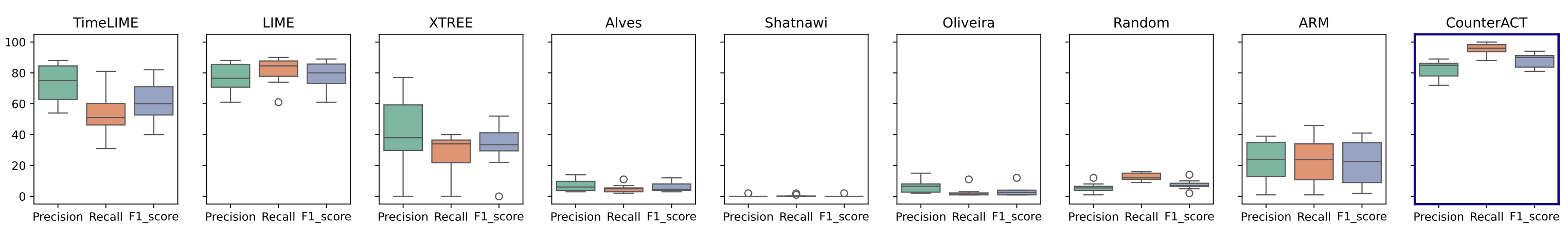}}
    \vspace{-4mm}
    \caption{Precision, Recall and F1-score distribution for each planner at the release level}
    \label{figure:pre_reca}
\vspace{-5mm}
\end{figure}

\textbf{The majority of recommendations proposed by CounterACT are implemented in the next release (high precision), 
and they are less likely to adopt changes that haven't been suggested (high recall).} 
In Figure~\ref{figure:pre_reca}, we display the distribution of precision, recall, and F1-scores of plans defined in \ref{sec:background}. The results reveal that 
CounterACT defect reduction planner achieves the highest precision score with 82.3\%, indicating that developers frequently implement CounterACT's plans.  Moreover, Figure~\ref{figure:pre_reca} presents the distribution of recall of 
plans, where a low value implies the presence of more unanticipated changes. CounterACT exhibits the highest recall 
(95.5\% on average) among all defect reduction planners, suggesting that it provides a variety of plans that are 
reflected in developers' actions. In fact, our data shows that developers are unlikely to implement changes that 
have not been recommended. 



\textbf{CounterACTs' plans exhibit the highest correlation between overlap and bug reductions in all projects}. 
Table~\ref{tab:NDPV} displays the average $S_{scaled}$ score (improvement score), where we can see that CounterACT attains the highest score across 
all projects : average $S_{scaled}$ of 91.33\%. Thus, when the overlap between the plan and actual changes is high, 
there tends to be an increase in code quality. Conversely, when the overlap is low, there tends to not be an increase in code
quality. 

Crucially, this does not prove that CounterACT \emph{fixed} more bugs than other planners. 
Making such a claim would require an \emph{interventionnal} study where we tell developers to follow the various plans.
Rather, a higher correlation means that the plans are less surprising since they overlap more with historical code 
changes that decrease the number of defects.

Furthermore, we conduct a statistical analysis of the empirical results, using the non-parametric paired Wilcoxon signed rank tests \cite{woolson2007wilcoxon} since the samples compared among planners are dependent. The null hypothesis of such tests is that CounterACT is as likely to yield a high-quality plan as any other planner. As shown in Table \ref{tab:overlap}, and Table \ref{tab:NDPV}, the p-value is less than 0.05 in all cases. This suggests that there is a statistically significant difference between the performances of CounterACT and the other approaches. Hence, there is strong evidence to reject the null hypothesis. Moreover, using the Wilcoxon signed rank tests across the different F1-scores, CounterACT demonstrates statistical significance compared to the other defect reduction planners with p-values consistently below the 0.05 threshold.

\begin{tcolorbox}
\vspace{-2mm}
At the release level, CounterACT plans are associated with greater defect reduction compared to competing approaches with an average $S_{scaled}$ score of 91.33\%. On average CounterACT plans overlap with actual developers' changes more than other approaches (95\%). Finally, CounterACT achieves the highest recall (95.5\%) and the best average F1-score (88.12\%), showing that developers would likely be able to implement its plans.
\vspace{-2mm}
\end{tcolorbox}
\input{tables/overlap_commit_IQR_release}

%% file: tables/overlap.tex
\begin{table*}[]
\vspace{-1mm}
    \begin{minipage}{.5\linewidth}
    \vspace{-3mm}
    \caption{Median overlap scores. A higher scores \\being preferable and indicated by dark shade}
    \vspace{-2mm}
    \centering
    \scalebox{0.55}{
        \begin{tabular}{|l|c|c|c|c|c|c|c|c|c|}
        \hline
        \textbf{} & \rotatebox{90}{TimeLIME} & \rotatebox{90}{LIME} & \rotatebox{90}{Shatnawi} & \rotatebox{90}{Alves} & \rotatebox{90}{Random} & \rotatebox{90}{Oliveira} & \rotatebox{90}{XTREE} & \rotatebox{90}{ARM$^{\mathrm{*}}$}& \rotatebox{90}{CounterACT} \\  \hline
        jedit & 70 & 60 & 50 & 30 & 17 & 35 & 75 & - &\cellcolor{gray!70}95\\
        camel1 & 90 & 85 & 65 & 60 & 15 & 62 & 80 & 34.27&\cellcolor{gray!70}95\\
        camel2 & 80 & 75 & 50 & 50 & 25 & 55 & 75 & 32.67&\cellcolor{gray!70}100\\
        log4j & 75 & 75 & 55 & 45 & 25 & 50 & 65 & - &\cellcolor{gray!70}95\\
        xalan & \cellcolor{gray!70}100 & \cellcolor{gray!70}100 & 75 & 80 & 65 & 85 & 75 &38.09 &\cellcolor{gray!70}100\\
        ant & 65 & 55 & 35 & 35 & 25 & 35 & 65 & - &\cellcolor{gray!70}90\\
        velocity & 90 & 85 & 70 & 55 & 20 & 75 & 80 &21.18 &\cellcolor{gray!70}95\\
        poi & 85 & 70 & 55 & 40 & 0 & 40 & 75 & - &\cellcolor{gray!70}95\\
        synapse & 67 & 70 & 40 & 40 & 10 & 42 & 62 & - &\cellcolor{gray!70}90\\
        
        \hline
        \rowcolor[HTML]{EFEFEF} 
        AVG & 79.17 & 75.56 & 55 & 48.33 & 22.55 & 53.33 & 72.22 &31.55 &\cellcolor{gray!70}95.00\\
        \rowcolor[HTML]{EFEFEF} 
        STD & 11.86 & 14.24 & 13.23 &  15.21 & 17.94 & 17.81 & 7.12 & 7.27&\cellcolor{gray!70}3.54 \\
        \rowcolor[HTML]{EFEFEF} 
        p-value$^{\mathrm{**}}$ & 0.01 & 0.01 & 0.003 & 0.003 &0.003  & 0.003 & 0.003 &0.001 &\cellcolor{gray!70}-\\
        \hline
        \multicolumn{10}{l}{$^{\mathrm{*}}$Action Rule Mining}\\
        \multicolumn{10}{l}{$^{\mathrm{**}}$Wilcoxon signed rank test to compare with CounterACT}
        
        \end{tabular}
    }
    \label{tab:overlap}
    \end{minipage}%
    \begin{minipage}{.5\linewidth}
        \vspace{-3mm}
      \caption{ Average $S_{scaled}$ score is presented. The top-performing planner is indicated by dark shade.}
      \vspace{-2mm}
        \centering
        \scalebox{0.55}{
        \begin{tabular}{|l|c|c|c|c|c|c|c|c|c|}
        \hline
        \textbf{} & \rotatebox{90}{TimeLIME} & \rotatebox{90}{LIME} & \rotatebox{90}{Shatnawi} & \rotatebox{90}{Alves} & \rotatebox{90}{Random} & \rotatebox{90}{Oliveira} & \rotatebox{90}{XTREE} & \rotatebox{90}{ARM$^{\mathrm{*}}$}& \rotatebox{90}{CounterACT} \\  \hline
        jedit   & 69 &  59 &  50 & 31 & 19 &  34 &  71 & - & \cellcolor{gray!70} 91  \\
          camel1  &  79 &  75 &  57 &  51 &  15 &  55 &  75 &33 &\cellcolor{gray!70} 92  \\
          camel2  &  68 &  56 &  37 & 31 & 13 &  42 &  71 & 29&\cellcolor{gray!70} 81  \\
          log4j    &  74 &  71 &  48 &  43 &  23 &  44 &  61 & - &\cellcolor{gray!70} 87  \\
          xalan    &  92 &  88 & 65 &  73 & 59 &  81 & 75 & 35 &\cellcolor{gray!70} 98 \\
          ant      &  82 &  72 & 59 &  46 & 40 &  60 & 69 &-  &\cellcolor{gray!70} 100\\
          velocity & 84 & 78 & 58 & 49 & 19 & 58 & 74 & 19 &\cellcolor{gray!70} 94\\
          poi      & 79 & 73 & 61 & 50 & 0 & 51 & 74 & - &\cellcolor{gray!70} 93 \\
          synapse  & 63 & 68 & 35 & 35 & 8 & 41 & 52 & - &\cellcolor{gray!70} 86 \\ 
          
        \hline
        \rowcolor[HTML]{EFEFEF} 
        AVG & 76.67 & 71.11  & 52.22 & 45.44 & 21.78 & 51.78 & 69.11 & 29 &\cellcolor{gray!70}91.33 \\
        \rowcolor[HTML]{EFEFEF} 
        STD & 9.06 &  9.60 & 10.57 & 13.02 & 17.75 & 13.96 & 7.77 & 7.11 &\cellcolor{gray!70} 5.96 \\
        \rowcolor[HTML]{EFEFEF} 
        p-value$^{\mathrm{**}}$ & 0.003 & 0.003  & 0.003 & 0.003 & 0.003 & 0.003 & 0.003 &0.001&\cellcolor{gray!70}-\\
        \hline
        \multicolumn{10}{l}{$^{\mathrm{*}}$Action Rule Mining}\\
        \multicolumn{10}{l}{$^{\mathrm{**}}$Wilcoxon signed rank test to compare with CounterACT}
        \end{tabular}   
        }
        \label{tab:NDPV}
    \end{minipage} 
    \vspace{-5mm}

\end{table*}

%% file: tables/overlap_commit_IQR_release.tex
\begin{table*}[]
    \vspace{-4mm}
    \begin{minipage}{.5\linewidth}
    
    \caption{Interquartile range (IQR) overlap \\ scores at release level: lower IQRs are \\preferred and are indicated by dark shade.}
    \vspace{-2mm}
    \scalebox{0.6}{
        \begin{tabular}{|l|c|c|c|c|c|c|c|c|c|}
\hline
\textbf{} & \rotatebox{90}{TimeLIME} & \rotatebox{90}{LIME} & \rotatebox{90}{Shatnawi} & \rotatebox{90}{Alves} & \rotatebox{90}{Random} & \rotatebox{90}{Oliveira} & \rotatebox{90}{XTREE} & \rotatebox{90}{ARM$^{\mathrm{*}}$}&\rotatebox{90}{CounterACT} \\  \hline
jedit      & 15.0    & 25.0 & 20.0  & 20.0 & 15.0     & 20.0     & 10.0  &-&\cellcolor{gray!50}10.0 \\
camel1     & 25.0    & 35.0 & 40.0  & 50.0 & 25.0     & 50.0     & 10.0  & 14.0&\cellcolor{gray!50}5.0 \\
camel2     & 20.0    & 35.0 & 30.0  & 45.0 & 30.0     & 50.0     & 6.25   & 15.0&\cellcolor{gray!50}5.0 \\
log4j      & 27.5    & 12.5 & 35.0  & 37.5 & 30.0     & 40.0     & 20.0  & -&\cellcolor{gray!50}10.0 \\
xalan      & 15.0    & 20.0 & 15.0  & 30.0 & 35.0     & 45.0     & \cellcolor{gray!40}0.0 & 5.0&\cellcolor{gray!30}5.0 \\
ant        & 25.0    & 27.5 & 20.0  & 15.0 & 17.5     & 20.0     & 20.0  & -&\cellcolor{gray!50}10.0 \\
velocity   & 10.0    & 30.0 & 25.0  & 35.0 & 15.0     & 50.0     & 5.0   & 15.0&\cellcolor{gray!50}0.0 \\
poi        & 15.0    & 15.0 & 15.0  & 20.0 & 10.0     & 20.0     & 10.0  & -&\cellcolor{gray!50}0.00 \\
synapse    & 38.75    & 25.0 & 30.0  & 38.75 & 15.0     & 35.0     & 23.75  & -&\cellcolor{gray!50}11.25 \\
\hline
\multicolumn{9}{l}{$^{\mathrm{*}}$Action Rule Mining}
\end{tabular}
    }
    \label{tab;IQR}
    \end{minipage}%
    \begin{minipage}{.5\linewidth}
      \caption{Median overlap scores (percentages) at the commit level. Higher scores are 
preferable and indicated by darker shading}
        \centering
        \scalebox{0.6}{
            \begin{tabular}{|l|c|c|c|c|c|c|c|c|c|c|}
            \hline
            & \rotatebox{90}{TimeLIME} & \rotatebox{90}{LIME} & \rotatebox{90}{Xtree} & \rotatebox{90}{Alves} & \rotatebox{90}{Shatnawi} & \rotatebox{90}{Oliveira} & \rotatebox{90}{ARM$^{\mathrm{*}}$}&\rotatebox{90}{CounterACT} \\ \hline
            kafka & 36.36 & 45.45 & 35.35 & 36.36 & 9.09 & 54.55 & 21.18&\cellcolor{gray!50}94.0\\
            activeMQ & 18.18 & 36.36 & 27.27 & 27.27 & 27.27 & 45.45 & 34.78& \cellcolor{gray!50}79.34\\
            nifi & 45.45 & 31.82 & 36.36 & 13.64 & 9.09 & 18.18 & 26.31&\cellcolor{gray!50}81.58\\
            zookeper & 45.45 & 27.27 & 24.24 & 13.64 & 9.09 & 18.18 & 28.09&\cellcolor{gray!50}88.95\\
            phoenix & 27.27 & 36.36 & 18.18 & 27.27 & 0.00 & 36.36 &31.81& \cellcolor{gray!50}74.47\\ 
            
            \hline
            \rowcolor[HTML]{EFEFEF} 
            AVG & 38.36 & 35.25 & 28.08 & 23.46 & 10.09 & 34.55 & 28.43&\cellcolor{gray!50}85.97\\ 
            \rowcolor[HTML]{EFEFEF} 
            STD & 13.68 &  \cellcolor{gray!50}6.09 &  7.33 & 9.04 &  9.598 & 15.26 & 5.21&\cellcolor{gray!40}6.4 \\ 
            \rowcolor[HTML]{EFEFEF} 
            p-value$^{\mathrm{**}}$ & 0.03 &  0.03 &  0.03 & 0.03 &  0.03 & 0.03 & 0.03&\cellcolor{gray!40}-\\ 
            \hline
            \multicolumn{9}{l}{$^{\mathrm{*}}$Action Rule Mining}\\
            \multicolumn{9}{l}{$^{\mathrm{**}}$Wilcoxon signed rank test to compare with CounterACT}

        \end{tabular}
        }
    \label{tab:overlap_commit}
    \end{minipage} 
    \vspace{-6mm}
\end{table*}

%% file: sections/RQ3.tex
\subsection*{\textbf{RQ3: How does CounterACT compare against competing defect reduction approaches at the commit level?}}
\setcounter{subsubsection}{0}


\indent

CounterACT is able to capture the relationship between bugs and software code metrics, even at a finer granularity (i.e., at the 
commit level). In Table~\ref{tab:overlap_commit}, 
we display how often the plans recommended by CounterACT overlap with code changes. CounterACT plans present the highest 
median overlap score (85.97\% on average), nearly twice that of the next highest competing approach TimeLIME (38.36\% on average) while the original ARM algorithm yields a median overlap score of 28.43\%.

\begin{figure*}[!h]
    \vspace{-4mm}
    \centering
    \centerline{\includegraphics[width=1\textwidth]
    {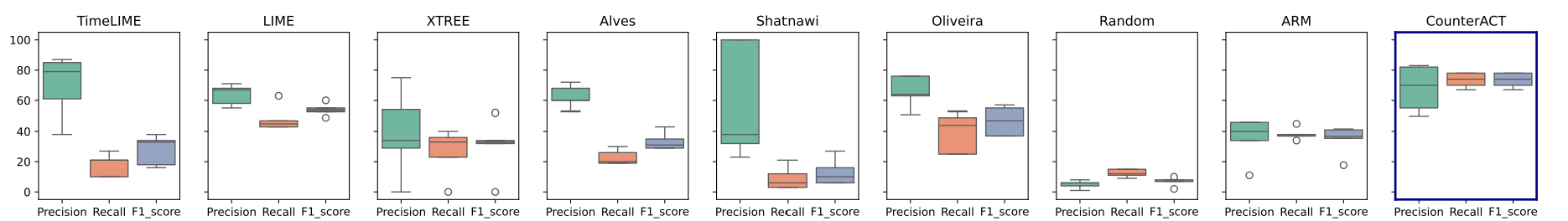}}
    \vspace{-3mm}
    \caption{Precision, Recall and F1-score distribution for each planner at the commit level}
    \label{figure:pre_reca_commit}
    \vspace{-3mm}
\end{figure*}

\textbf{The majority of recommendations proposed by CounterACT are adopted in future commits 
(high precision), and they are less likely to adopt changes that haven’t been suggested (high recall) at a finer granularity}. 
In Figure~\ref{figure:pre_reca_commit}, we display the precision and recall of plans for various defect reduction planners. 
The results reveal that CounterACT achieves not only the highest scores in both precision and recall with respective median values 
of 78\% and 75\% but also the most optimal precision-recall trade-off (average F1-score 72.83\%), indicating that developers 
can potentially implement the majority of CounterACT's recommendations. In fact, developers are less likely to implement changes 
that have not been recommended. 

In summary, CounterACT produces stable plans by keeping a consistent performance at both release and commit granularities 
(median overlap score of 95\% and 85.97\% respectively), outperforming other defect reduction planners in terms of providing 
plans that align more closely with developers' choices. This suggests that counterfactual analysis is an efficient method for 
defect reduction planning, as it increases transparency by eliminating the need for black-box models. 

Moreover, we study the statistical significance of the overlap scores and the F1-scores at the commit level. Similarly to RQ2, we use the Wilcoxon signed rank test. As shown in Table \ref{tab:overlap_commit}, the p-value is less than 0.05 in every case. This suggests that there is a statistically significant difference between the overlap performance of CounterACT and the other approaches. Moreover, using Wilcoxon signed rank tests across the F1-scores, CounterACT demonstrates statistical significance compared to the other planners having p-values consistently below the 0.05 threshold.

\begin{tcolorbox}
\vspace{-2mm}
CounterACT produces more feasible plans than competing defect reduction approaches, as evidenced by the overlap score observed actions at commit -level (median overlap of 85.97\% Table~\ref{tab:overlap_commit}).
In addition, CounterACT has the highest \textbf{Precision} and \textbf{Recall} trade-off among all the competing methods (average \textbf{F1-score} 72.83\%). 
\vspace{-2mm}
\end{tcolorbox}

%% file: sections/4_discussion.tex
\section{Discussion \& implications}
The aim of this section is to discuss the implications of our approach by exploring the potential effectiveness and feasibility of our proposed planning approach.


\begin{figure*}[!h]
\vspace{-10mm}
\centering
\centerline{\includegraphics[width=0.93\textwidth]{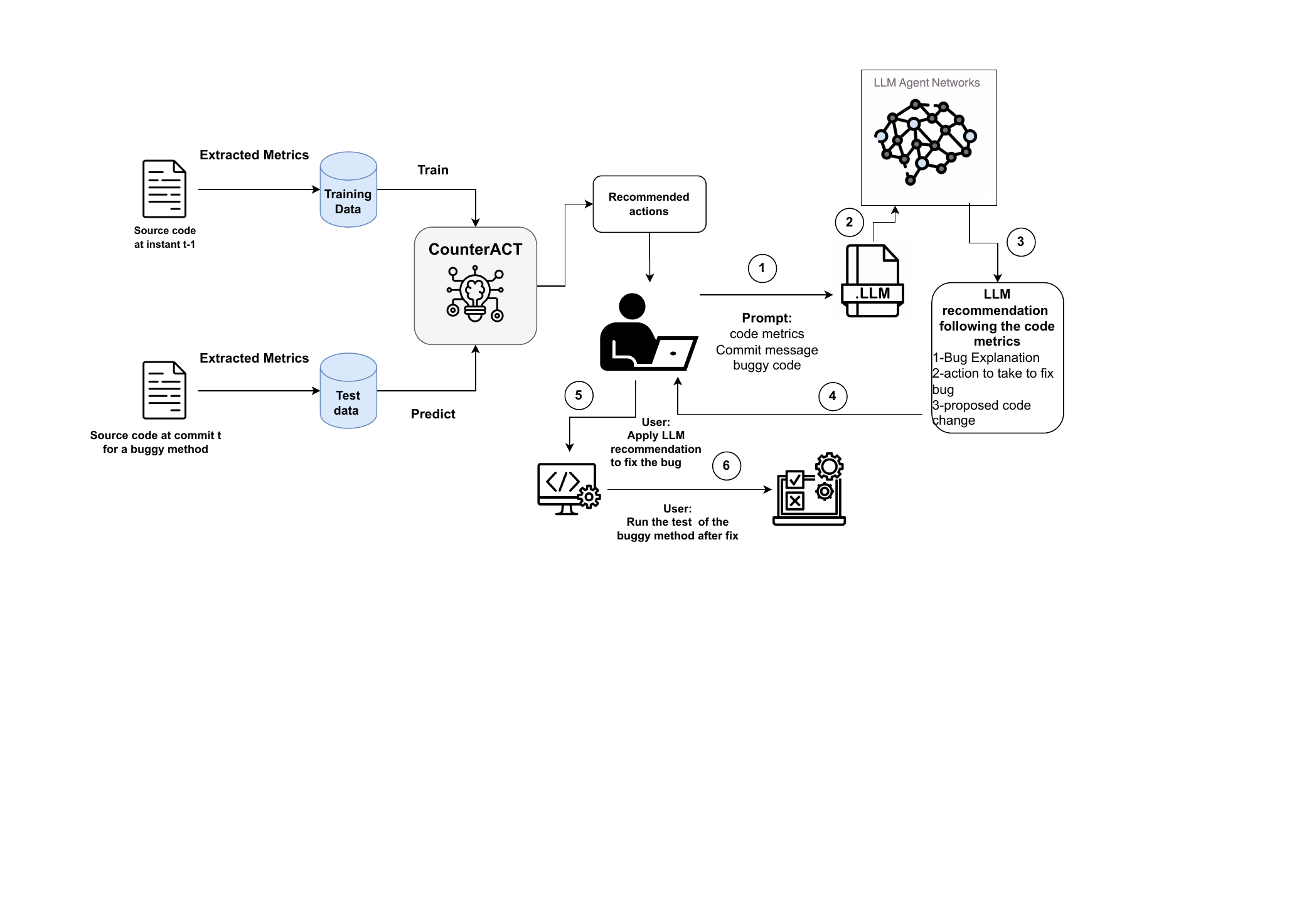}}
\vspace{-40mm}
\caption{Overview: Incorporate the LLM with CounterACT to generate actionable recommendations
for developers}
\vspace{-3mm}
\label{figure:overviewLLM}
\end{figure*}

\textbf{Feasibility of our plans for developers:} Our research questions do not directly evaluate whether it is feasible for developers to apply CounterACT's plans. To remedy this limitation, we aim to provide some verification of their practical feasibility at the commit-level. To do so, we introduce an additional layer of support: the use of a LLMs (Large language models). Integrating an LLM with our generated plans aims to provide more actionable recommendations for developers.

To evaluate the efficacy of the LLM in aiding developers to act on CounterACT plans, we conduct experiments using  CodeLlama~\cite{rozière2023code}. In figure \ref{figure:overviewLLM}
we show the process of using LLM along with CounterACT to transform our generated plans into actions developers can implement.

We conduct experiments using the CodeLlama2~\cite{rozière2023code} LLM 34b trained by Meta. The model is fed the following information: the commit message, the code requiring a fix, and, importantly, the guidelines provided by CounterACT. These guidelines specifically articulate how the code metrics should be altered. In fact, our prompt followed a predefined template ~\cite{facebookresearch_2023_llama-recipes} (an example can be found in the reproduction package). Our experiments focused on the Kafka project, targeting 40 randomly selected bugs. For validation, we relied on the codeLlama2 model's guidance on code and ran test cases on the buggy method to ascertain whether the bugs had indeed been rectified. Our observations were as follows:


\indent \textbf{Effectiveness with guidance:} When provided with our plans' guidance, the LLM generates actionable guidance for 40 randomly selected bugs. This guidance takes the form of bug explanations and a set of actions to take to fix the bug coupled with relevant code snippets in the illustrated example~\ref{lst:pre_pa}. In 28 out of the 40 bugs when following the LLM recommendation, the bug is fixed and the test cases pass. This highlights the potential of combining a LLM with strategic planning, leading to tangible and actionable solutions.

\indent \textbf{Effectiveness without guidance:} When the vanilla LLM was only provided with the buggy code and commit message, its suggestions were less useful. Without guidance, the code snippets generated by the LLM lacked coherence and resulted in more failed test cases. Indeed, 26 out of 40 of these suggestions resulted in failed test cases. 14 of those failing cases were prevented when using CounterACT plans, showing that our generated plans can guide the LLM in providing more useful suggestions.

\begin{figure}[!h]
\vspace{-4mm}
\centering
\centerline{\includegraphics[width=0.85\textwidth]{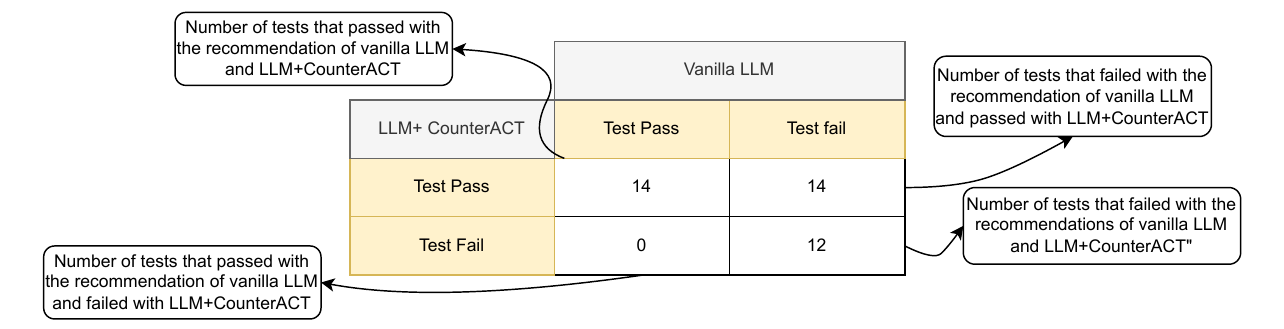}}
\vspace{-2mm}
\caption{Contingency table to Compare test outcomes for the vanilla LLM proposed bug fix VS the LLM with CounterACT
guidance.}
\vspace{-5mm}
\label{figure:LLM_contigency_Table}
\end{figure}
\input{sections/LLM_code_exemple} 

Moreover, we study the statistical significance of the LLM results, we use the McNemar statistical test \cite{PemburySmith2020} which aims at detecting differences between paired samples of two boolean variables. We organize the samples in a 2x2 contingency table as shown below in Figure \ref{figure:LLM_contigency_Table}. The null hypothesis of the McNemar test is that the distributions of unit tests passing with and without guidance are identical. Under this assumption, the off-diagonal terms in the table should be evenly distributed. Here, we see that no unit test passed without guidance and failed with guidance. Additionally, 14 unit tests passed with guidance and failed without guidance. It is worth noting that there are cases where the LLM often encounters difficulties despite CounterACT’s guidance because the bug is specifically related to interactions with other classes or dependencies on different sections of the codebase. Since we do not provide such context to the LLM, no bug fix can be found in those instances. Despite a lack of context in certain cases, our results, when compared to LLM bug fixes without guidance, still yield a McNemar statistical test p-value of 0.000183 which is much less than 0.05. This provides evidence to reject the null hypothesis of no change, indicating that there is a statistically significant difference when using the LLM with guidance compared to not using it. 
While it is challenging to ascertain the exposure of LLMs to the project’s fixing patches due to a lack of training details, we show that our enhancement to the LLM (through CounterACT) does indeed generally improve results since the vanilla LLM code edits failed the tests in 26 out of 40 experiments despite the risk of data leakage, all while demonstrating statistical significance.

The recommendations of LLMs are stochastic by design, different runs with the same prompt can induce different code fixes. This
introduces an additional source of variability which may invalidate our previous observations. To address this issue, each inference of the LLM was run 10 times \cite{stvr.1486}. More specifically, for each unit test with/without CounterACT's guidance, we compute 10  LLM inferences and record the rates of success of the test. Since such rates are no longer boolean variables, the McNemar test is not applicable. Thus, we use a Repeated-Measure-ANOVA (RM-ANOVA) test \cite{rutherford2011anova}. The test yields a significant p-value of $1.0\times10^{-4}\%$. 

%% file: sections/LLM_code_exemple.tex
\begin{figure}[!ht]
\captionsetup{type=lstlisting}

\begin{sublstlisting}{\linewidth}
\vspace{-3 mm}
\begin{lstlisting}[language=Java]
// repository : apache/kafka
// commit id:c83719f3d8799ec35cf9af95fd93c97b910df570
 public boolean hasRoomFor(long timestamp, ByteBuffer key, ByteBuffer value) {
        if (isFull())
            return false;
        ...
            - recordSize = DefaultRecord.sizeInBytes(nextOffsetDelta, timestampDelta, key, 
            value, Record.EMPTY_HEADERS);
            + recordSize = DefaultRecord.sizeInBytes(nextOffsetDelta, timestampDelta, key, value, headers);
\end{lstlisting}
\vspace{-2mm}
\caption{Buggy code, the line with (-) represents the bug. The line with (+) represents the real bug fix from developers}
\end{sublstlisting}
\vspace{-2mm}
\begin{sublstlisting}{\linewidth}
\begin{lstlisting}[language=Java]
//The bug is caused by the hasRoomFor method not taking into account the size of the record headers when determining whether there is enough room in the buffer for another record. As a result, it may allow records to be appended that exceed the maximum allowed size, leading to unexpected behavior or errors.
//To fix this issue, I would suggest changing the implementation as follows:
public boolean hasRoomFor(long timestamp, ByteBuffer key, ByteBuffer value, Header[] headers) {
        if (isFull())
            return false;
            ....
        recordSize = DefaultRecord.sizeInBytes(nextOffsetDelta, timestampDelta, key, value, headers);
    }
    return recordSize <= maxRecordSize;
\end{lstlisting}
\vspace{-2mm}
\caption{Proposed bug fix with the generated plan from \counteract}
\end{sublstlisting}
\vspace{-3mm}

\begin{sublstlisting}{\linewidth}
\begin{lstlisting}[language=Java]
//I would recommend adding some error handling or logging to show that the record could not be added due to lack of space. For example, you could add a log message like "Not enough space to add record" before returning null
public FutureRecordMetadata tryAppend(long timestamp, byte[] key, byte[] value, Header[] headers, Callback callback, long now) {
    if (recordsBuilder.hasRoomFor(timestamp, key, value)) {
        ...
    } else {
        logger.warn("Not enough space to add record");
        return null;
\end{lstlisting}
\vspace{-3mm}
\caption{Proposed fix without the plan generated from \counteract}
\vspace{-2mm}
\end{sublstlisting}
\vspace{-5mm}

\caption{The code snippet shown in (a) triggers a bug in the Kafka project at line 7. In (b) we show the bug fix recommendation of codeLlama when given the proposed plan \textbf{"NUMPAR:(0.995, 1.005)=>(1.025, 2.005) and don't change the rest metrics"}. The LLM has provided good recommendations and fixed the method \textbf{hasRoomFor}. As in (c), we show the code snippet of the recommendation of the vanilla LLM. We see that the model ignores the bug and provides a logging statement to just track the bug. 
(You can find the prompt in our replication package.) }
\label{lst:pre_pa}
\vspace{-3mm}
\end{figure}

%% file: sections/5_threatsToValidity.tex
\section{Threats to Validity} 
\label{sec:threats}

\paragraph{Internal Validity}

Our study might be subject to potential alternative explanations for the observed outcomes. Although the use of alternative software metrics is not considered a direct threat, a pertinent concern arises when questioning the effectiveness of our chosen metrics in reducing defects. It is possible that our metrics are merely correlated with defect reduction, rather than being causally linked to it. This raises the possibility of other, unexplored factors or explanations that could be responsible for the observed decrease in defects. To mitigate this concern, we rely on metrics that have been widely used in prior works. While exploring the causal relationships of metrics is beyond the scope of this work, future work should involve examining additional metrics and factors.
Moreover, the question remains whether the suggested plans can be credited for defect reduction or suppression. We believe that even if the plans do not guarantee defect reduction, they are correlated with defect reduction and can therefore provide a step toward improvement.

In our paper, the plans generated by CounterACT aim to identify factors commonly associated with improved software quality outcomes. For instance, a suggestion to “change the number of parameters” is for a file-level instance and the software metrics are specified to the object (e.g., class, method, interface). While this is still relatively general, it can prompt a review of methods that are overly complex or hard to maintain, guiding developers toward considering refactoring or redesigning for better modularity and readability. We recognize the importance of generating immediately actionable recommendations for practitioners. Thus we see value in how LLMs code edits were improved with CounterACT’s guidance. We perceive this as a fruitful direction for future research to further refine the approach with fault-localization techniques and larger datasets.

\paragraph{External Validity}

In our work, we focused on datasets consisting only of Apache projects. This can constrain the generalizability of our findings across diverse software development environments. However, we used Apache projects to maintain consistency with related work \cite{peng2021defect,krishna2020prediction}. Using the same context allows for direct comparison, thereby enhancing the validity and comparability of our findings. Future work should aim to extend the analysis to more diverse datasets to validate and expand our findings' applicability.


We implemented CounterACT on a selected set of software projects, which may limit the generalizability of our results to other datasets. To counter this limitation, we chose a range of real-world, open-source software applications for our analysis. Nevertheless, conducting further replication studies in alternative ecosystems would be beneficial for validating our findings.

While the release-level data employed states that it contains logs of past defects, we cannot guarantee the usefulness of overlap as an argument for claiming that we have captured specific bug fixes made by developers. Nonetheless, similarly to prior studies, overlap is an essential sanity check to ensure that the recommendations made by CounterACT are realistic. To complement overlap, we also measure the precision and recall. Finally, we also evaluate the approach at the commit level where labels are more specific to defects, thus minimizing confusion.

On the commit level, we use the public dataset provided by \cite{Trautsch_PROMISE_2021_Defect_2021} which was processed using the base SZZ algorithm. We acknowledge that recent updates to the SZZ algorithm (e.g., RA-SZZ \cite{8330225}) could very likely improve the results of our study, however by relying on an existing dataset to improve replicability and comparisons we were constrained to the choices made by that dataset. Future work could likely benefit from creating new datasets using more recent SZZ versions.


%% file: sections/conclusion.tex
\vspace{-2mm}
\section{Conclusion}
\label{sec:conclusion}
To the best of our knowledge, this study is the first to explore counterfactual-based planning using
action rules mining for defect reduction. In this paper, we assess CounterACT, our novel approach for generating defect reduction plans, against state-of-the-art defect reduction planners on both release level and commit level. Results show that the plans are:
\vspace{-1mm}
\begin{itemize}

    \item Plausible: They are aligned with developers' actions in future releases or commits. CounterACT demonstrates better \textbf{Overlap}, \textbf{Precision}, \textbf{Recall}, and \textbf{F1-score} compared to competing approaches. Even at a finer granularity, the performance of our approach remains consistent. This shows that counterfactual analysis is effective at generating defect reduction plans and capturing relationships between code metrics even if the change is small. 
    \item Actionable: Supporting \texttt{LLM} with CounterACT planning results in feasible recommendations. This has been evidenced by the fact that out of 40 bugs selected using this method, 28 have been successfully fixed and passed their respective test cases.
\end{itemize}

This research demonstrates that white-box approaches can provide effective defect-reduction plans
that are competitive with state-of-the-art black-box approaches while offering a different view of what can be done in explainable software analytics.



\vspace{-2mm}
\section{Data Availability}
\label{sec:dataAvailability}
Our data, and the scripts necessary to replicate our work is available, under an open license, using the following link: \url{https://github.com/yukikh1234/counterACT_Defect_Reduction_Planning}

%% file: sample-manuscript.bbl

\begin{thebibliography}{68}


\ifx \showCODEN    \undefined \def \showCODEN     #1{\unskip}     \fi
\ifx \showDOI      \undefined \def \showDOI       #1{#1}\fi
\ifx \showISBNx    \undefined \def \showISBNx     #1{\unskip}     \fi
\ifx \showISBNxiii \undefined \def \showISBNxiii  #1{\unskip}     \fi
\ifx \showISSN     \undefined \def \showISSN      #1{\unskip}     \fi
\ifx \showLCCN     \undefined \def \showLCCN      #1{\unskip}     \fi
\ifx \shownote     \undefined \def \shownote      #1{#1}          \fi
\ifx \showarticletitle \undefined \def \showarticletitle #1{#1}   \fi
\ifx \showURL      \undefined \def \showURL       {\relax}        \fi
\providecommand\bibfield[2]{#2}
\providecommand\bibinfo[2]{#2}
\providecommand\natexlab[1]{#1}
\providecommand\showeprint[2][]{arXiv:#2}

\bibitem[Agrawal and Menzies(2017)]%
        {DBLP:journals/corr/AgrawalM17}
\bibfield{author}{\bibinfo{person}{Amritanshu Agrawal} {and} \bibinfo{person}{Tim Menzies}.} \bibinfo{year}{2017}\natexlab{}.
\newblock \showarticletitle{"Better Data" is Better than "Better Data Miners" (Benefits of Tuning {SMOTE} for Defect Prediction)}.
\newblock \bibinfo{journal}{\emph{CoRR}}  \bibinfo{volume}{abs/1705.03697} (\bibinfo{year}{2017}).
\newblock
\showeprint[arXiv]{1705.03697}
\urldef\tempurl%
\url{http://arxiv.org/abs/1705.03697}
\showURL{%
\tempurl}


\bibitem[Alves et~al\mbox{.}(2010)]%
        {alves2010deriving}
\bibfield{author}{\bibinfo{person}{Tiago~L Alves}, \bibinfo{person}{Christiaan Ypma}, {and} \bibinfo{person}{Joost Visser}.} \bibinfo{year}{2010}\natexlab{}.
\newblock \showarticletitle{Deriving metric thresholds from benchmark data}. In \bibinfo{booktitle}{\emph{2010 IEEE international conference on software maintenance}}. IEEE, \bibinfo{pages}{1--10}.
\newblock


\bibitem[Arcuri and Briand(2014)]%
        {stvr.1486}
\bibfield{author}{\bibinfo{person}{Andrea Arcuri} {and} \bibinfo{person}{Lionel Briand}.} \bibinfo{year}{2014}\natexlab{}.
\newblock \showarticletitle{A Hitchhiker's guide to statistical tests for assessing randomized algorithms in software engineering}.
\newblock \bibinfo{journal}{\emph{Software Testing, Verification and Reliability}} \bibinfo{volume}{24}, \bibinfo{number}{3} (\bibinfo{year}{2014}), \bibinfo{pages}{219--250}.
\newblock
\urldef\tempurl%
\url{https://doi.org/10.1002/stvr.1486}
\showDOI{\tempurl}
\showeprint{https://onlinelibrary.wiley.com/doi/pdf/10.1002/stvr.1486}


\bibitem[Boehm and Basili(2001)]%
        {Boehm}
\bibfield{author}{\bibinfo{person}{Barry Boehm} {and} \bibinfo{person}{Victor Basili}.} \bibinfo{year}{2001}\natexlab{}.
\newblock \showarticletitle{Software Defect Reduction Top 10 List}.
\newblock \bibinfo{journal}{\emph{IEEE Computer}}  \bibinfo{volume}{34} (\bibinfo{date}{01} \bibinfo{year}{2001}), \bibinfo{pages}{135--137}.
\newblock
\showISBNx{978-3-540-24547-6}
\urldef\tempurl%
\url{https://doi.org/10.1007/3-540-27662-9_26}
\showDOI{\tempurl}


\bibitem[Chawla et~al\mbox{.}(2002)]%
        {smote}
\bibfield{author}{\bibinfo{person}{Nitesh Chawla}, \bibinfo{person}{Kevin Bowyer}, \bibinfo{person}{Lawrence Hall}, {and} \bibinfo{person}{W. Kegelmeyer}.} \bibinfo{year}{2002}\natexlab{}.
\newblock \showarticletitle{SMOTE: Synthetic Minority Over-sampling Technique}.
\newblock \bibinfo{journal}{\emph{J. Artif. Intell. Res. (JAIR)}}  \bibinfo{volume}{16} (\bibinfo{date}{06} \bibinfo{year}{2002}), \bibinfo{pages}{321--357}.
\newblock
\urldef\tempurl%
\url{https://doi.org/10.1613/jair.953}
\showDOI{\tempurl}


\bibitem[Chidamber and Kemerer(1994)]%
        {295895}
\bibfield{author}{\bibinfo{person}{S.R. Chidamber} {and} \bibinfo{person}{C.F. Kemerer}.} \bibinfo{year}{1994}\natexlab{}.
\newblock \showarticletitle{A metrics suite for object oriented design}.
\newblock \bibinfo{journal}{\emph{IEEE Transactions on Software Engineering}} \bibinfo{volume}{20}, \bibinfo{number}{6} (\bibinfo{year}{1994}), \bibinfo{pages}{476--493}.
\newblock
\urldef\tempurl%
\url{https://doi.org/10.1109/32.295895}
\showDOI{\tempurl}


\bibitem[Couto et~al\mbox{.}(2014)]%
        {COUTO201424}
\bibfield{author}{\bibinfo{person}{Cesar Couto}, \bibinfo{person}{Pedro Pires}, \bibinfo{person}{Marco~Tulio Valente}, \bibinfo{person}{Roberto~S. Bigonha}, {and} \bibinfo{person}{Nicolas Anquetil}.} \bibinfo{year}{2014}\natexlab{}.
\newblock \showarticletitle{Predicting software defects with causality tests}.
\newblock \bibinfo{journal}{\emph{Journal of Systems and Software}}  \bibinfo{volume}{93} (\bibinfo{year}{2014}), \bibinfo{pages}{24--41}.
\newblock
\showISSN{0164-1212}
\urldef\tempurl%
\url{https://doi.org/10.1016/j.jss.2014.01.033}
\showDOI{\tempurl}


\bibitem[Dam et~al\mbox{.}(2018)]%
        {DBLP:journals/corr/abs-1802-00603}
\bibfield{author}{\bibinfo{person}{Hoa~Khanh Dam}, \bibinfo{person}{Truyen Tran}, {and} \bibinfo{person}{Aditya Ghose}.} \bibinfo{year}{2018}\natexlab{}.
\newblock \showarticletitle{Explainable Software Analytics}.
\newblock \bibinfo{journal}{\emph{CoRR}}  \bibinfo{volume}{abs/1802.00603} (\bibinfo{year}{2018}).
\newblock
\showeprint[arXiv]{1802.00603}
\urldef\tempurl%
\url{http://arxiv.org/abs/1802.00603}
\showURL{%
\tempurl}


\bibitem[D'Ambros et~al\mbox{.}(2010)]%
        {AmbrosAS}
\bibfield{author}{\bibinfo{person}{Marco D'Ambros}, \bibinfo{person}{Michele Lanza}, {and} \bibinfo{person}{Romain Robbes}.} \bibinfo{year}{2010}\natexlab{}.
\newblock \showarticletitle{An extensive comparison of bug prediction approaches}.
\newblock \bibinfo{journal}{\emph{Proceedings - International Conference on Software Engineering}}, \bibinfo{pages}{31--41}.
\newblock
\urldef\tempurl%
\url{https://doi.org/10.1109/MSR.2010.5463279}
\showDOI{\tempurl}


\bibitem[Dardzinska(2012)]%
        {ActionRuleMining}
\bibfield{author}{\bibinfo{person}{Agnieszka Dardzinska}.} \bibinfo{year}{2012}\natexlab{}.
\newblock \bibinfo{booktitle}{\emph{Action rules mining}}. Vol.~\bibinfo{volume}{468}.
\newblock \bibinfo{publisher}{Springer Berlin, Heidelberg}.
\newblock


\bibitem[Fayyad and Irani(1993)]%
        {Fayyad1993MultiIntervalDO}
\bibfield{author}{\bibinfo{person}{Usama~M. Fayyad} {and} \bibinfo{person}{Keki~B. Irani}.} \bibinfo{year}{1993}\natexlab{}.
\newblock \showarticletitle{Multi-Interval Discretization of Continuous-Valued Attributes for Classification Learning}. In \bibinfo{booktitle}{\emph{International Joint Conference on Artificial Intelligence}}.
\newblock
\urldef\tempurl%
\url{https://api.semanticscholar.org/CorpusID:18718011}
\showURL{%
\tempurl}


\bibitem[Frederick~P(1995)]%
        {hawking1988}
\bibfield{author}{\bibinfo{person}{Brooks~Jr. Frederick~P}.} \bibinfo{year}{1995}\natexlab{}.
\newblock \bibinfo{booktitle}{\emph{The Mythical Man-Month: Essays on Software Engineering}}.
\newblock \bibinfo{publisher}{Pearson Education}.
\newblock


\bibitem[Gezici and Tarhan(2022)]%
        {9919490}
\bibfield{author}{\bibinfo{person}{Bahar Gezici} {and} \bibinfo{person}{Ayça~Kolukisa Tarhan}.} \bibinfo{year}{2022}\natexlab{}.
\newblock \showarticletitle{Explainable AI for Software Defect Prediction with Gradient Boosting Classifier}. In \bibinfo{booktitle}{\emph{2022 7th International Conference on Computer Science and Engineering (UBMK)}}. \bibinfo{pages}{1--6}.
\newblock
\urldef\tempurl%
\url{https://doi.org/10.1109/UBMK55850.2022.9919490}
\showDOI{\tempurl}


\bibitem[Hall et~al\mbox{.}(2012)]%
        {6035727}
\bibfield{author}{\bibinfo{person}{Tracy Hall}, \bibinfo{person}{Sarah Beecham}, \bibinfo{person}{David Bowes}, \bibinfo{person}{David Gray}, {and} \bibinfo{person}{Steve Counsell}.} \bibinfo{year}{2012}\natexlab{}.
\newblock \showarticletitle{A Systematic Literature Review on Fault Prediction Performance in Software Engineering}.
\newblock \bibinfo{journal}{\emph{IEEE Transactions on Software Engineering}} \bibinfo{volume}{38}, \bibinfo{number}{6} (\bibinfo{year}{2012}), \bibinfo{pages}{1276--1304}.
\newblock
\urldef\tempurl%
\url{https://doi.org/10.1109/TSE.2011.103}
\showDOI{\tempurl}


\bibitem[Herbold et~al\mbox{.}(2020)]%
        {DBLP:journals/corr/abs-2003-05357}
\bibfield{author}{\bibinfo{person}{Steffen Herbold}, \bibinfo{person}{Alexander Trautsch}, {and} \bibinfo{person}{Fabian Trautsch}.} \bibinfo{year}{2020}\natexlab{}.
\newblock \showarticletitle{On the Feasibility of Automated Issue Type Prediction}.
\newblock \bibinfo{journal}{\emph{CoRR}}  \bibinfo{volume}{abs/2003.05357} (\bibinfo{year}{2020}).
\newblock
\showeprint[arXiv]{2003.05357}
\urldef\tempurl%
\url{https://arxiv.org/abs/2003.05357}
\showURL{%
\tempurl}


\bibitem[Hihn and Menzies(2015)]%
        {7426628}
\bibfield{author}{\bibinfo{person}{Jaitus Hihn} {and} \bibinfo{person}{Tim Menzies}.} \bibinfo{year}{2015}\natexlab{}.
\newblock \showarticletitle{Data Mining Methods and Cost Estimation Models: Why is it So Hard to Infuse New Ideas?}. In \bibinfo{booktitle}{\emph{2015 30th IEEE/ACM International Conference on Automated Software Engineering Workshop (ASEW)}}. \bibinfo{pages}{5--9}.
\newblock
\urldef\tempurl%
\url{https://doi.org/10.1109/ASEW.2015.27}
\showDOI{\tempurl}


\bibitem[Jiarpakdee et~al\mbox{.}(2020)]%
        {Jiarpakdee}
\bibfield{author}{\bibinfo{person}{Jirayus Jiarpakdee}, \bibinfo{person}{Chakkrit Tantithamthavorn}, \bibinfo{person}{Hoa Dam}, {and} \bibinfo{person}{John Grundy}.} \bibinfo{year}{2020}\natexlab{}.
\newblock \showarticletitle{An Empirical Study of Model-Agnostic Techniques for Defect Prediction Models}.
\newblock \bibinfo{journal}{\emph{IEEE Transactions on Software Engineering}}  \bibinfo{volume}{PP} (\bibinfo{date}{03} \bibinfo{year}{2020}), \bibinfo{pages}{1--1}.
\newblock
\urldef\tempurl%
\url{https://doi.org/10.1109/TSE.2020.2982385}
\showDOI{\tempurl}


\bibitem[Jiarpakdee et~al\mbox{.}(2018)]%
        {autospearman}
\bibfield{author}{\bibinfo{person}{Jirayus Jiarpakdee}, \bibinfo{person}{Chakkrit Tantithamthavorn}, {and} \bibinfo{person}{Christoph Treude}.} \bibinfo{year}{2018}\natexlab{}.
\newblock \showarticletitle{AutoSpearman: Automatically Mitigating Correlated Metrics for Interpreting Defect Models}.
\newblock  (\bibinfo{date}{06} \bibinfo{year}{2018}).
\newblock


\bibitem[Jovanoski and Lavra{\v{c}}(2001)]%
        {10.1007/3-540-45329-6_8}
\bibfield{author}{\bibinfo{person}{Viktor Jovanoski} {and} \bibinfo{person}{Nada Lavra{\v{c}}}.} \bibinfo{year}{2001}\natexlab{}.
\newblock \showarticletitle{Classification Rule Learning with APRIORI-C}. In \bibinfo{booktitle}{\emph{Progress in Artificial Intelligence}}, \bibfield{editor}{\bibinfo{person}{Pavel Brazdil} {and} \bibinfo{person}{Al{\'i}pio Jorge}} (Eds.). \bibinfo{publisher}{Springer Berlin Heidelberg}, \bibinfo{address}{Berlin, Heidelberg}, \bibinfo{pages}{44--51}.
\newblock


\bibitem[Jureczko(2011)]%
        {article}
\bibfield{author}{\bibinfo{person}{Marian Jureczko}.} \bibinfo{year}{2011}\natexlab{}.
\newblock \showarticletitle{Significance of Different Software Metrics in Defect Prediction}.
\newblock \bibinfo{journal}{\emph{Software Engineering: An International Journal}}  \bibinfo{volume}{1} (\bibinfo{date}{01} \bibinfo{year}{2011}), \bibinfo{pages}{86--95}.
\newblock


\bibitem[Jureczko and Madeyski(2010)]%
        {10.1145/1868328.1868342}
\bibfield{author}{\bibinfo{person}{Marian Jureczko} {and} \bibinfo{person}{Lech Madeyski}.} \bibinfo{year}{2010}\natexlab{}.
\newblock \showarticletitle{Towards Identifying Software Project Clusters with Regard to Defect Prediction} \emph{(\bibinfo{series}{PROMISE '10})}. \bibinfo{publisher}{Association for Computing Machinery}, \bibinfo{address}{New York, NY, USA}, Article \bibinfo{articleno}{9}, \bibinfo{numpages}{10}~pages.
\newblock
\showISBNx{9781450304047}
\urldef\tempurl%
\url{https://doi.org/10.1145/1868328.1868342}
\showDOI{\tempurl}


\bibitem[Kamei et~al\mbox{.}(2013a)]%
        {article3}
\bibfield{author}{\bibinfo{person}{Yasutaka Kamei}, \bibinfo{person}{Emad Shihab}, \bibinfo{person}{Bram Adams}, \bibinfo{person}{Ahmed~E. Hassan}, \bibinfo{person}{Audris Mockus}, \bibinfo{person}{Anand Sinha}, {and} \bibinfo{person}{Naoyasu Ubayashi}.} \bibinfo{year}{2013}\natexlab{a}.
\newblock \showarticletitle{A Large-Scale Empirical Study of Just-in-Time Quality Assurance}.
\newblock \bibinfo{journal}{\emph{Software Engineering, IEEE Transactions on}}  \bibinfo{volume}{39} (\bibinfo{date}{06} \bibinfo{year}{2013}), \bibinfo{pages}{757--773}.
\newblock
\urldef\tempurl%
\url{https://doi.org/10.1109/TSE.2012.70}
\showDOI{\tempurl}


\bibitem[Kamei et~al\mbox{.}(2013b)]%
        {6341763}
\bibfield{author}{\bibinfo{person}{Yasutaka Kamei}, \bibinfo{person}{Emad Shihab}, \bibinfo{person}{Bram Adams}, \bibinfo{person}{Ahmed~E. Hassan}, \bibinfo{person}{Audris Mockus}, \bibinfo{person}{Anand Sinha}, {and} \bibinfo{person}{Naoyasu Ubayashi}.} \bibinfo{year}{2013}\natexlab{b}.
\newblock \showarticletitle{A large-scale empirical study of just-in-time quality assurance}.
\newblock \bibinfo{journal}{\emph{IEEE Transactions on Software Engineering}} \bibinfo{volume}{39}, \bibinfo{number}{6} (\bibinfo{year}{2013}), \bibinfo{pages}{757--773}.
\newblock
\urldef\tempurl%
\url{https://doi.org/10.1109/TSE.2012.70}
\showDOI{\tempurl}


\bibitem[Kampenes et~al\mbox{.}(2007)]%
        {Kampenes}
\bibfield{author}{\bibinfo{person}{Vigdis Kampenes}, \bibinfo{person}{Tore Dybå}, \bibinfo{person}{Jo Hannay}, {and} \bibinfo{person}{Dag Sjøberg}.} \bibinfo{year}{2007}\natexlab{}.
\newblock \showarticletitle{A systematic review of effect size in software engineering experiments}.
\newblock \bibinfo{journal}{\emph{Information and Software Technology}}  \bibinfo{volume}{49} (\bibinfo{date}{11} \bibinfo{year}{2007}), \bibinfo{pages}{1073--1086}.
\newblock
\urldef\tempurl%
\url{https://doi.org/10.1016/j.infsof.2007.02.015}
\showDOI{\tempurl}


\bibitem[Kim et~al\mbox{.}(2007)]%
        {10.1109/ICSE.2007.66}
\bibfield{author}{\bibinfo{person}{Sunghun Kim}, \bibinfo{person}{Thomas Zimmermann}, \bibinfo{person}{E.~James Whitehead~Jr.}, {and} \bibinfo{person}{Andreas Zeller}.} \bibinfo{year}{2007}\natexlab{}.
\newblock \showarticletitle{Predicting Faults from Cached History}. In \bibinfo{booktitle}{\emph{Proceedings of the 29th International Conference on Software Engineering}} \emph{(\bibinfo{series}{ICSE '07})}. \bibinfo{publisher}{IEEE Computer Society}, \bibinfo{address}{USA}, \bibinfo{pages}{489–498}.
\newblock
\showISBNx{0769528287}
\urldef\tempurl%
\url{https://doi.org/10.1109/ICSE.2007.66}
\showDOI{\tempurl}


\bibitem[Krishna and Menzies(2017)]%
        {Krishna2017FromPT}
\bibfield{author}{\bibinfo{person}{Rahul Krishna} {and} \bibinfo{person}{Tim Menzies}.} \bibinfo{year}{2017}\natexlab{}.
\newblock \showarticletitle{From Prediction to Planning: Improving Software Quality with BELLTREE}.
\newblock \bibinfo{journal}{\emph{arXiv: Software Engineering}} (\bibinfo{year}{2017}).
\newblock


\bibitem[Krishna and Menzies(2020)]%
        {krishna2020prediction}
\bibfield{author}{\bibinfo{person}{Rahul Krishna} {and} \bibinfo{person}{Tim Menzies}.} \bibinfo{year}{2020}\natexlab{}.
\newblock \showarticletitle{From prediction to planning: Improving software quality with BELLTREE}.
\newblock \bibinfo{journal}{\emph{Empir. Softw. Eng.}}  \bibinfo{volume}{25} (\bibinfo{year}{2020}), \bibinfo{pages}{3468--3500}.
\newblock


\bibitem[Kumar and Kumar(2018)]%
        {Kumar2018UpliftM}
\bibfield{author}{\bibinfo{person}{Akshay Kumar} {and} \bibinfo{person}{Rishabh Kumar}.} \bibinfo{year}{2018}\natexlab{}.
\newblock \showarticletitle{Uplift Modeling : Predicting incremental gains}.
\newblock
\urldef\tempurl%
\url{https://api.semanticscholar.org/CorpusID:113397046}
\showURL{%
\tempurl}


\bibitem[Lee and Lee(2023)]%
        {10123486}
\bibfield{author}{\bibinfo{person}{Gichan Lee} {and} \bibinfo{person}{Scott Uk-Jin Lee}.} \bibinfo{year}{2023}\natexlab{}.
\newblock \showarticletitle{An Empirical Comparison of Model-Agnostic Techniques for Defect Prediction Models}. In \bibinfo{booktitle}{\emph{2023 IEEE International Conference on Software Analysis, Evolution and Reengineering (SANER)}}. \bibinfo{pages}{179--189}.
\newblock
\urldef\tempurl%
\url{https://doi.org/10.1109/SANER56733.2023.00026}
\showDOI{\tempurl}


\bibitem[Lema{\^i}tre et~al\mbox{.}(2016)]%
        {Lematre2016ImbalancedlearnAP}
\bibfield{author}{\bibinfo{person}{Guillaume Lema{\^i}tre}, \bibinfo{person}{Fernando Nogueira}, {and} \bibinfo{person}{Christos~K. Aridas}.} \bibinfo{year}{2016}\natexlab{}.
\newblock \showarticletitle{Imbalanced-learn: A Python Toolbox to Tackle the Curse of Imbalanced Datasets in Machine Learning}.
\newblock \bibinfo{journal}{\emph{ArXiv}}  \bibinfo{volume}{abs/1609.06570} (\bibinfo{year}{2016}).
\newblock
\urldef\tempurl%
\url{https://api.semanticscholar.org/CorpusID:1426815}
\showURL{%
\tempurl}


\bibitem[Lundberg and Lee(2017)]%
        {DBLP:journals/corr/LundbergL17}
\bibfield{author}{\bibinfo{person}{Scott~M. Lundberg} {and} \bibinfo{person}{Su{-}In Lee}.} \bibinfo{year}{2017}\natexlab{}.
\newblock \showarticletitle{A unified approach to interpreting model predictions}.
\newblock \bibinfo{journal}{\emph{CoRR}}  \bibinfo{volume}{abs/1705.07874} (\bibinfo{year}{2017}).
\newblock
\showeprint[arXiv]{1705.07874}
\urldef\tempurl%
\url{http://arxiv.org/abs/1705.07874}
\showURL{%
\tempurl}


\bibitem[McCabe(1976)]%
        {1702388}
\bibfield{author}{\bibinfo{person}{T.J. McCabe}.} \bibinfo{year}{1976}\natexlab{}.
\newblock \showarticletitle{A Complexity Measure}.
\newblock \bibinfo{journal}{\emph{IEEE Transactions on Software Engineering}} \bibinfo{volume}{SE-2}, \bibinfo{number}{4} (\bibinfo{year}{1976}), \bibinfo{pages}{308--320}.
\newblock
\urldef\tempurl%
\url{https://doi.org/10.1109/TSE.1976.233837}
\showDOI{\tempurl}


\bibitem[Mende and Koschke(2010)]%
        {inproceedings1}
\bibfield{author}{\bibinfo{person}{Thilo Mende} {and} \bibinfo{person}{Rainer Koschke}.} \bibinfo{year}{2010}\natexlab{}.
\newblock \showarticletitle{Effort-Aware Defect Prediction Models}.
\newblock \bibinfo{journal}{\emph{European Conference on Software Maintenance and Reengineering}}, \bibinfo{pages}{107--116}.
\newblock
\urldef\tempurl%
\url{https://doi.org/10.1109/CSMR.2010.18}
\showDOI{\tempurl}


\bibitem[Menzies et~al\mbox{.}(2007a)]%
        {Menzies121}
\bibfield{author}{\bibinfo{person}{Tim Menzies}, \bibinfo{person}{Jeremy Greenwald}, {and} \bibinfo{person}{Art Frank}.} \bibinfo{year}{2007}\natexlab{a}.
\newblock \showarticletitle{Data Mining Static Code Attributes to Learn Defect Predictors.}
\newblock \bibinfo{journal}{\emph{IEEE Trans. Software Eng.}}  \bibinfo{volume}{33} (\bibinfo{date}{01} \bibinfo{year}{2007}), \bibinfo{pages}{2--13}.
\newblock
\urldef\tempurl%
\url{https://doi.org/10.1109/TSE.2007.10}
\showDOI{\tempurl}


\bibitem[Menzies et~al\mbox{.}(2007b)]%
        {4027145}
\bibfield{author}{\bibinfo{person}{Tim Menzies}, \bibinfo{person}{Jeremy Greenwald}, {and} \bibinfo{person}{Art Frank}.} \bibinfo{year}{2007}\natexlab{b}.
\newblock \showarticletitle{Data Mining Static Code Attributes to Learn Defect Predictors}.
\newblock \bibinfo{journal}{\emph{IEEE Transactions on Software Engineering}} \bibinfo{volume}{33}, \bibinfo{number}{1} (\bibinfo{year}{2007}), \bibinfo{pages}{2--13}.
\newblock
\urldef\tempurl%
\url{https://doi.org/10.1109/TSE.2007.256941}
\showDOI{\tempurl}


\bibitem[Mobini and Banihashemi(2016)]%
        {Mobini}
\bibfield{author}{\bibinfo{person}{Mina Mobini} {and} \bibinfo{person}{Sepideh Banihashemi}.} \bibinfo{year}{2016}\natexlab{}.
\newblock \showarticletitle{Analysis of Defect Prediction by using Object Oriented Metrics.}
\newblock  (\bibinfo{date}{04} \bibinfo{year}{2016}).
\newblock
\urldef\tempurl%
\url{https://doi.org/10.13140/RG.2.1.4075.8166}
\showDOI{\tempurl}


\bibitem[Mothilal et~al\mbox{.}(2020)]%
        {mothilal2020explaining}
\bibfield{author}{\bibinfo{person}{Ramaravind~K Mothilal}, \bibinfo{person}{Amit Sharma}, {and} \bibinfo{person}{Chenhao Tan}.} \bibinfo{year}{2020}\natexlab{}.
\newblock \showarticletitle{Explaining machine learning classifiers through diverse counterfactual explanations}. In \bibinfo{booktitle}{\emph{Proceedings of the 2020 conference on fairness, accountability, and transparency}}. \bibinfo{pages}{607--617}.
\newblock


\bibitem[Nagappan and Ball(2007)]%
        {inproceedings2}
\bibfield{author}{\bibinfo{person}{Nachiappan Nagappan} {and} \bibinfo{person}{Thomas Ball}.} \bibinfo{year}{2007}\natexlab{}.
\newblock \showarticletitle{Using Software Dependencies and Churn Metrics to Predict Field Failures: An Empirical Case Study}.
\newblock \bibinfo{journal}{\emph{Proceedings - 1st International Symposium on Empirical Software Engineering and Measurement, ESEM 2007}}, \bibinfo{pages}{364--373}.
\newblock
\showISBNx{978-0-7695-2886-1}
\urldef\tempurl%
\url{https://doi.org/10.1109/ESEM.2007.13}
\showDOI{\tempurl}


\bibitem[Nagappan et~al\mbox{.}(2010)]%
        {NagappanQWQ}
\bibfield{author}{\bibinfo{person}{Nachiappan Nagappan}, \bibinfo{person}{Andreas Zeller}, \bibinfo{person}{Thomas Zimmermann}, \bibinfo{person}{Kim Herzig}, {and} \bibinfo{person}{Brendan Murphy}.} \bibinfo{year}{2010}\natexlab{}.
\newblock \showarticletitle{Change Bursts as Defect Predictors}.
\newblock \bibinfo{journal}{\emph{Proceedings - International Symposium on Software Reliability Engineering, ISSRE}}.
\newblock
\urldef\tempurl%
\url{https://doi.org/10.1109/ISSRE.2010.25}
\showDOI{\tempurl}


\bibitem[Neto et~al\mbox{.}(2018)]%
        {8330225}
\bibfield{author}{\bibinfo{person}{Edmilson~Campos Neto}, \bibinfo{person}{Daniel~Alencar da Costa}, {and} \bibinfo{person}{Uirá Kulesza}.} \bibinfo{year}{2018}\natexlab{}.
\newblock \showarticletitle{The impact of refactoring changes on the SZZ algorithm: An empirical study}. In \bibinfo{booktitle}{\emph{2018 IEEE 25th International Conference on Software Analysis, Evolution and Reengineering (SANER)}}. \bibinfo{pages}{380--390}.
\newblock
\urldef\tempurl%
\url{https://doi.org/10.1109/SANER.2018.8330225}
\showDOI{\tempurl}


\bibitem[Oliveira et~al\mbox{.}(2014)]%
        {oliveira2014extracting}
\bibfield{author}{\bibinfo{person}{Paloma Oliveira}, \bibinfo{person}{Marco~Tulio Valente}, {and} \bibinfo{person}{Fernando~Paim Lima}.} \bibinfo{year}{2014}\natexlab{}.
\newblock \showarticletitle{Extracting relative thresholds for source code metrics}. In \bibinfo{booktitle}{\emph{2014 Software Evolution Week-IEEE Conference on Software Maintenance, Reengineering, and Reverse Engineering (CSMR-WCRE)}}. IEEE, \bibinfo{pages}{254--263}.
\newblock


\bibitem[Parmentier and Vidal(2021)]%
        {parmentier2021optimal}
\bibfield{author}{\bibinfo{person}{Axel Parmentier} {and} \bibinfo{person}{Thibaut Vidal}.} \bibinfo{year}{2021}\natexlab{}.
\newblock \showarticletitle{Optimal counterfactual explanations in tree ensembles}. In \bibinfo{booktitle}{\emph{International Conference on Machine Learning}}. PMLR, \bibinfo{pages}{8422--8431}.
\newblock


\bibitem[Pascarella et~al\mbox{.}(2019)]%
        {PASCARELLA201922}
\bibfield{author}{\bibinfo{person}{Luca Pascarella}, \bibinfo{person}{Fabio Palomba}, {and} \bibinfo{person}{Alberto Bacchelli}.} \bibinfo{year}{2019}\natexlab{}.
\newblock \showarticletitle{Fine-grained just-in-time defect prediction}.
\newblock \bibinfo{journal}{\emph{Journal of Systems and Software}}  \bibinfo{volume}{150} (\bibinfo{year}{2019}), \bibinfo{pages}{22--36}.
\newblock
\showISSN{0164-1212}
\urldef\tempurl%
\url{https://doi.org/10.1016/j.jss.2018.12.001}
\showDOI{\tempurl}


\bibitem[Pembury~Smith and Ruxton(2020)]%
        {PemburySmith2020}
\bibfield{author}{\bibinfo{person}{Matilda Q.~R. Pembury~Smith} {and} \bibinfo{person}{Graeme~D. Ruxton}.} \bibinfo{year}{2020}\natexlab{}.
\newblock \showarticletitle{Effective use of the McNemar test}.
\newblock \bibinfo{journal}{\emph{Behavioral Ecology and Sociobiology}} \bibinfo{volume}{74}, \bibinfo{number}{11} (\bibinfo{year}{2020}), \bibinfo{pages}{133}.
\newblock
\showISSN{1432-0762}
\urldef\tempurl%
\url{https://doi.org/10.1007/s00265-020-02916-y}
\showDOI{\tempurl}


\bibitem[Peng and Menzies(2021)]%
        {peng2021defect}
\bibfield{author}{\bibinfo{person}{Kewen Peng} {and} \bibinfo{person}{Tim Menzies}.} \bibinfo{year}{2021}\natexlab{}.
\newblock \showarticletitle{Defect reduction planning (using timelime)}.
\newblock \bibinfo{journal}{\emph{IEEE Transactions on Software Engineering}} \bibinfo{volume}{48}, \bibinfo{number}{7} (\bibinfo{year}{2021}), \bibinfo{pages}{2510--2525}.
\newblock


\bibitem[Pornprasit and Tantithamthavorn(2021)]%
        {DBLP:journals/corr/abs-2103-07068}
\bibfield{author}{\bibinfo{person}{Chanathip Pornprasit} {and} \bibinfo{person}{Chakkrit Tantithamthavorn}.} \bibinfo{year}{2021}\natexlab{}.
\newblock \showarticletitle{JITLine: {A} Simpler, Better, Faster, Finer-grained Just-In-Time Defect Prediction}.
\newblock \bibinfo{journal}{\emph{CoRR}}  \bibinfo{volume}{abs/2103.07068} (\bibinfo{year}{2021}).
\newblock
\showeprint[arXiv]{2103.07068}
\urldef\tempurl%
\url{https://arxiv.org/abs/2103.07068}
\showURL{%
\tempurl}


\bibitem[Pornprasit et~al\mbox{.}(2021)]%
        {9678763}
\bibfield{author}{\bibinfo{person}{Chanathip Pornprasit}, \bibinfo{person}{Chakkrit Tantithamthavorn}, \bibinfo{person}{Jirayus Jiarpakdee}, \bibinfo{person}{Michael Fu}, {and} \bibinfo{person}{Patanamon Thongtanunam}.} \bibinfo{year}{2021}\natexlab{}.
\newblock \showarticletitle{PyExplainer: Explaining the Predictions of Just-In-Time Defect Models}. In \bibinfo{booktitle}{\emph{2021 36th IEEE/ACM International Conference on Automated Software Engineering (ASE)}}. \bibinfo{pages}{407--418}.
\newblock
\urldef\tempurl%
\url{https://doi.org/10.1109/ASE51524.2021.9678763}
\showDOI{\tempurl}


\bibitem[Rajapaksha et~al\mbox{.}(2021)]%
        {sqaplanner}
\bibfield{author}{\bibinfo{person}{Dilini Rajapaksha}, \bibinfo{person}{Chakkrit Tantithamthavorn}, \bibinfo{person}{Jirayus Jiarpakdee}, \bibinfo{person}{Christoph Bergmeir}, \bibinfo{person}{John Grundy}, {and} \bibinfo{person}{Wray Buntine}.} \bibinfo{year}{2021}\natexlab{}.
\newblock \showarticletitle{SQAPlanner: Generating Data-Informed Software Quality Improvement Plans}.
\newblock \bibinfo{journal}{\emph{IEEE Transactions on Software Engineering}}  \bibinfo{volume}{PP} (\bibinfo{date}{04} \bibinfo{year}{2021}), \bibinfo{pages}{1--1}.
\newblock
\urldef\tempurl%
\url{https://doi.org/10.1109/TSE.2021.3070559}
\showDOI{\tempurl}


\bibitem[Research(2023)]%
        {facebookresearch_2023_llama-recipes}
\bibfield{author}{\bibinfo{person}{Facebook Research}.} \bibinfo{year}{2023}\natexlab{}.
\newblock \bibinfo{title}{llama-recipes}.
\newblock
\newblock
\urldef\tempurl%
\url{https://github.com/facebookresearch/llama-recipes/tree/main}
\showURL{%
\tempurl}
\newblock
\shownote{Accessed: 2023-06-28}.


\bibitem[Ribeiro et~al\mbox{.}(2016)]%
        {ribeiro2016should}
\bibfield{author}{\bibinfo{person}{Marco~Tulio Ribeiro}, \bibinfo{person}{Sameer Singh}, {and} \bibinfo{person}{Carlos Guestrin}.} \bibinfo{year}{2016}\natexlab{}.
\newblock \showarticletitle{" Why should i trust you?" Explaining the predictions of any classifier}. In \bibinfo{booktitle}{\emph{Proceedings of the 22nd ACM SIGKDD international conference on knowledge discovery and data mining}}. \bibinfo{pages}{1135--1144}.
\newblock


\bibitem[Rozière et~al\mbox{.}(2023)]%
        {rozière2023code}
\bibfield{author}{\bibinfo{person}{Baptiste Rozière}, \bibinfo{person}{Jonas Gehring}, \bibinfo{person}{Fabian Gloeckle}, \bibinfo{person}{Sten Sootla}, \bibinfo{person}{Itai Gat}, \bibinfo{person}{Xiaoqing~Ellen Tan}, \bibinfo{person}{Yossi Adi}, \bibinfo{person}{Jingyu Liu}, \bibinfo{person}{Tal Remez}, \bibinfo{person}{Jérémy Rapin}, \bibinfo{person}{Artyom Kozhevnikov}, \bibinfo{person}{Ivan Evtimov}, \bibinfo{person}{Joanna Bitton}, \bibinfo{person}{Manish Bhatt}, \bibinfo{person}{Cristian~Canton Ferrer}, \bibinfo{person}{Aaron Grattafiori}, \bibinfo{person}{Wenhan Xiong}, \bibinfo{person}{Alexandre Défossez}, \bibinfo{person}{Jade Copet}, \bibinfo{person}{Faisal Azhar}, \bibinfo{person}{Hugo Touvron}, \bibinfo{person}{Louis Martin}, \bibinfo{person}{Nicolas Usunier}, \bibinfo{person}{Thomas Scialom}, {and} \bibinfo{person}{Gabriel Synnaeve}.} \bibinfo{year}{2023}\natexlab{}.
\newblock \bibinfo{title}{Code Llama: Open Foundation Models for Code}.
\newblock
\newblock
\showeprint[arxiv]{2308.12950}~[cs.CL]


\bibitem[Rudin(2019)]%
        {rudin2019stop}
\bibfield{author}{\bibinfo{person}{Cynthia Rudin}.} \bibinfo{year}{2019}\natexlab{}.
\newblock \bibinfo{title}{Stop Explaining Black Box Machine Learning Models for High Stakes Decisions and Use Interpretable Models Instead}.
\newblock
\newblock
\showeprint[arxiv]{1811.10154}~[stat.ML]


\bibitem[Russell(2019)]%
        {russell2019efficient}
\bibfield{author}{\bibinfo{person}{Chris Russell}.} \bibinfo{year}{2019}\natexlab{}.
\newblock \showarticletitle{Efficient search for diverse coherent explanations}. In \bibinfo{booktitle}{\emph{Proceedings of the Conference on Fairness, Accountability, and Transparency}}. \bibinfo{pages}{20--28}.
\newblock


\bibitem[Rutherford(2011)]%
        {rutherford2011anova}
\bibfield{author}{\bibinfo{person}{Andrew Rutherford}.} \bibinfo{year}{2011}\natexlab{}.
\newblock \bibinfo{booktitle}{\emph{ANOVA and ANCOVA: a GLM approach}}.
\newblock \bibinfo{publisher}{John Wiley \& Sons}.
\newblock


\bibitem[Shatnawi(2010)]%
        {shatnawi2010quantitative}
\bibfield{author}{\bibinfo{person}{Raed Shatnawi}.} \bibinfo{year}{2010}\natexlab{}.
\newblock \showarticletitle{A quantitative investigation of the acceptable risk levels of object-oriented metrics in open-source systems}.
\newblock \bibinfo{journal}{\emph{IEEE Transactions on software engineering}} \bibinfo{volume}{36}, \bibinfo{number}{2} (\bibinfo{year}{2010}), \bibinfo{pages}{216--225}.
\newblock


\bibitem[Shin et~al\mbox{.}(2021a)]%
        {shin2021explainable}
\bibfield{author}{\bibinfo{person}{Jiho Shin}, \bibinfo{person}{Reem Aleithan}, \bibinfo{person}{Jaechang Nam}, \bibinfo{person}{Junjie Wang}, {and} \bibinfo{person}{Song Wang}.} \bibinfo{year}{2021}\natexlab{a}.
\newblock \bibinfo{title}{Explainable Software Defect Prediction: Are We There Yet?}
\newblock
\newblock
\showeprint[arxiv]{2111.10901}~[cs.SE]


\bibitem[Shin et~al\mbox{.}(2021b)]%
        {DBLP:journals/corr/abs-2111-10901}
\bibfield{author}{\bibinfo{person}{Jiho Shin}, \bibinfo{person}{Reem Aleithan}, \bibinfo{person}{Jaechang Nam}, \bibinfo{person}{Junjie Wang}, {and} \bibinfo{person}{Song Wang}.} \bibinfo{year}{2021}\natexlab{b}.
\newblock \showarticletitle{Explainable Software Defect Prediction: Are We There Yet?}
\newblock \bibinfo{journal}{\emph{CoRR}}  \bibinfo{volume}{abs/2111.10901} (\bibinfo{year}{2021}).
\newblock
\showeprint[arXiv]{2111.10901}
\urldef\tempurl%
\url{https://arxiv.org/abs/2111.10901}
\showURL{%
\tempurl}


\bibitem[Sliwerski et~al\mbox{.}(2005)]%
        {inproceedings}
\bibfield{author}{\bibinfo{person}{Jacek Sliwerski}, \bibinfo{person}{Thomas Zimmermann}, {and} \bibinfo{person}{Andreas Zeller}.} \bibinfo{year}{2005}\natexlab{}.
\newblock \showarticletitle{When do changes induce fixes?}
\newblock \bibinfo{journal}{\emph{ACM Sigsoft Software Engineering Notes}}  \bibinfo{volume}{30}.
\newblock
\urldef\tempurl%
\url{https://doi.org/10.1145/1082983.1083147}
\showDOI{\tempurl}


\bibitem[Sykora and Kliegr(2020)]%
        {Sykora2020ActionRC}
\bibfield{author}{\bibinfo{person}{Lukas Sykora} {and} \bibinfo{person}{Tom{\'a}{\'s} Kliegr}.} \bibinfo{year}{2020}\natexlab{}.
\newblock \showarticletitle{Action Rules: Counterfactual Explanations in Python}. In \bibinfo{booktitle}{\emph{RuleML+RR}}.
\newblock


\bibitem[Tantithamthavorn et~al\mbox{.}(2020)]%
        {8494821}
\bibfield{author}{\bibinfo{person}{Chakkrit Tantithamthavorn}, \bibinfo{person}{Ahmed~E. Hassan}, {and} \bibinfo{person}{Kenichi Matsumoto}.} \bibinfo{year}{2020}\natexlab{}.
\newblock \showarticletitle{The Impact of Class Rebalancing Techniques on the Performance and Interpretation of Defect Prediction Models}.
\newblock \bibinfo{journal}{\emph{IEEE Transactions on Software Engineering}} \bibinfo{volume}{46}, \bibinfo{number}{11} (\bibinfo{year}{2020}), \bibinfo{pages}{1200--1219}.
\newblock
\urldef\tempurl%
\url{https://doi.org/10.1109/TSE.2018.2876537}
\showDOI{\tempurl}


\bibitem[Tantithamthavorn et~al\mbox{.}(2021)]%
        {tantithamthavorn2021actionable}
\bibfield{author}{\bibinfo{person}{Chakkrit Tantithamthavorn}, \bibinfo{person}{Jirayus Jiarpakdee}, {and} \bibinfo{person}{John Grundy}.} \bibinfo{year}{2021}\natexlab{}.
\newblock \showarticletitle{Actionable analytics: Stop telling me what it is; please tell me what to do}.
\newblock \bibinfo{journal}{\emph{IEEE Software}} \bibinfo{volume}{38}, \bibinfo{number}{4} (\bibinfo{year}{2021}), \bibinfo{pages}{115--120}.
\newblock


\bibitem[Tantithamthavorn et~al\mbox{.}(2019)]%
        {8263202}
\bibfield{author}{\bibinfo{person}{Chakkrit Tantithamthavorn}, \bibinfo{person}{Shane McIntosh}, \bibinfo{person}{Ahmed~E. Hassan}, {and} \bibinfo{person}{Kenichi Matsumoto}.} \bibinfo{year}{2019}\natexlab{}.
\newblock \showarticletitle{The Impact of Automated Parameter Optimization on Defect Prediction Models}.
\newblock \bibinfo{journal}{\emph{IEEE Transactions on Software Engineering}} \bibinfo{volume}{45}, \bibinfo{number}{7} (\bibinfo{year}{2019}), \bibinfo{pages}{683--711}.
\newblock
\urldef\tempurl%
\url{https://doi.org/10.1109/TSE.2018.2794977}
\showDOI{\tempurl}


\bibitem[Trautsch et~al\mbox{.}(2021)]%
        {DBLP:journals/corr/abs-2109-03544}
\bibfield{author}{\bibinfo{person}{Alexander Trautsch}, \bibinfo{person}{Johannes Erbel}, \bibinfo{person}{Steffen Herbold}, {and} \bibinfo{person}{Jens Grabowski}.} \bibinfo{year}{2021}\natexlab{}.
\newblock \showarticletitle{On the differences between quality increasing and other changes in open source Java projects}.
\newblock \bibinfo{journal}{\emph{CoRR}}  \bibinfo{volume}{abs/2109.03544} (\bibinfo{year}{2021}).
\newblock
\showeprint[arXiv]{2109.03544}
\urldef\tempurl%
\url{https://arxiv.org/abs/2109.03544}
\showURL{%
\tempurl}


\bibitem[Trautsch and Herbold(2021)]%
        {Trautsch_PROMISE_2021_Defect_2021}
\bibfield{author}{\bibinfo{person}{Alexander Trautsch} {and} \bibinfo{person}{Steffen Herbold}.} \bibinfo{year}{2021}\natexlab{}.
\newblock \bibinfo{booktitle}{\emph{{PROMISE 2021 Defect Prediction Challenge Data}}}.
\newblock
\urldef\tempurl%
\url{https://doi.org/10.5281/zenodo.8370181}
\showDOI{\tempurl}


\bibitem[Wachter et~al\mbox{.}(2017)]%
        {wachter2017counterfactual}
\bibfield{author}{\bibinfo{person}{Sandra Wachter}, \bibinfo{person}{Brent Mittelstadt}, {and} \bibinfo{person}{Chris Russell}.} \bibinfo{year}{2017}\natexlab{}.
\newblock \showarticletitle{Counterfactual explanations without opening the black box: Automated decisions and the GDPR}.
\newblock \bibinfo{journal}{\emph{Harv. JL \& Tech.}}  \bibinfo{volume}{31} (\bibinfo{year}{2017}), \bibinfo{pages}{841}.
\newblock


\bibitem[Wahono(2015)]%
        {Wahono}
\bibfield{author}{\bibinfo{person}{Romi Wahono}.} \bibinfo{year}{2015}\natexlab{}.
\newblock \showarticletitle{A Systematic Literature Review of Software Defect Prediction: Research Trends, Datasets, Methods and Frameworks}.
\newblock \bibinfo{journal}{\emph{Journal of Software Engineering}}  \bibinfo{volume}{1} (\bibinfo{date}{05} \bibinfo{year}{2015}).
\newblock


\bibitem[Woolson(2007)]%
        {woolson2007wilcoxon}
\bibfield{author}{\bibinfo{person}{Robert~F Woolson}.} \bibinfo{year}{2007}\natexlab{}.
\newblock \showarticletitle{Wilcoxon signed-rank test}.
\newblock \bibinfo{journal}{\emph{Wiley encyclopedia of clinical trials}} (\bibinfo{year}{2007}), \bibinfo{pages}{1--3}.
\newblock


\bibitem[Yang et~al\mbox{.}(2016)]%
        {10.1145/2950290.2950353}
\bibfield{author}{\bibinfo{person}{Yibiao Yang}, \bibinfo{person}{Yuming Zhou}, \bibinfo{person}{Jinping Liu}, \bibinfo{person}{Yangyang Zhao}, \bibinfo{person}{Hongmin Lu}, \bibinfo{person}{Lei Xu}, \bibinfo{person}{Baowen Xu}, {and} \bibinfo{person}{Hareton Leung}.} \bibinfo{year}{2016}\natexlab{}.
\newblock \showarticletitle{Effort-Aware Just-in-Time Defect Prediction: Simple Unsupervised Models Could Be Better than Supervised Models}. In \bibinfo{booktitle}{\emph{Proceedings of the 2016 24th ACM SIGSOFT International Symposium on Foundations of Software Engineering}} (Seattle, WA, USA) \emph{(\bibinfo{series}{FSE 2016})}. \bibinfo{publisher}{Association for Computing Machinery}, \bibinfo{address}{New York, NY, USA}, \bibinfo{pages}{157–168}.
\newblock
\showISBNx{9781450342186}
\urldef\tempurl%
\url{https://doi.org/10.1145/2950290.2950353}
\showDOI{\tempurl}


\end{thebibliography}
